\documentclass[preprint,a4paper,12pt,3p]{elsarticle} 
\usepackage{graphicx}
\usepackage{graphics}
\usepackage[mathscr]{eucal}
\usepackage{amssymb}
\usepackage{amsmath}
\usepackage{tabls} 
\usepackage{multirow}
\usepackage{color}

\usepackage{listings}

\newcommand{\Lag}{\mathcal{L}}
\newcommand{\Nge}{\mathcal{N}}
\newcommand{\Mass}{\mathcal{M}}

\newcommand{\prog}{{\tt CNNDecays}}
\def\rpv{$R_p \hspace{-1em}/\;\:\hspace{0.2em}$}

\begin{document}

\begin{frontmatter}

\title{On-shell renormalization of neutralino and chargino mass matrices
in $R$-parity violating models - Correlation between LSP decays
and neutrino mixing angles revisited}

\author[wue]{Stefan Liebler}
\ead{sliebler@physik.uni-wuerzburg.de}

\author[wue]{Werner Porod}
\ead{porod@physik.uni-wuerzburg.de}

\address[wue]{
Institut f\"ur Theoretische Physik und Astrophysik, Universit\"at W\"urzburg,\\
D-97074  W\"urzburg, Germany
}

\begin{abstract}
We present an on-shell renormalization scheme for the neutralino and chargino mass
matrices for two $R$-parity violating models, namely bilinear $R$-parity violation
and the $\mu\nu$SSM with one right-handed neutrino superfield.
We discuss in both models how to obtain neutrino masses and mixing
angles correctly as well as differences to the existing $\overline{\text{DR}}$
calculations. Both models predict correlations between
neutrino mixing angles and ratios of $R$-parity violating decays such as
$\tilde{\chi}_1^0\rightarrow l^\pm W^\mp$ which can be tested
for example at the LHC. We point out that there are
cases where one has to include the complete NLO corrections 
for the decays $\tilde{\chi}_1^0\rightarrow l^\pm W^\mp$ to predict
reliably the correlations between neutrino- and collider-physics.
Moreover, we discuss briefly the implementation of these decays
in the public program \prog. 
\end{abstract}

\end{frontmatter}

\tableofcontents

\section{Introduction}
\label{sec:introduction}

The Minimal Supersymmetric extension of the Standard Model (MSSM)
\cite{Martin:1997ns} is based on conserved $R$-parity, where
$R$-parity ($R_p$) \cite{Farrar:1978xj} is defined as
$R_p = (-1)^{3B+L+2s}$ and originally guaranteed the stability
of the proton in supersymmetric models \cite{Weinberg:1981wj}.
It implies on the one hand a stable lightest supersymmetric
particle (LSP) and predicts on the other hand zero neutrino masses
as in the Standard Model (SM).
As neutrino oscillation experiments have shown at least two neutrinos
have non-zero mass \cite{Fukuda:1998mi}
requiring an extension to explain neutrino data\footnote{A
recent analysis combining the newest precision results from various
 experiments like
KamLAND \cite{Gando:2010aa}, Super-K \cite{Cravens:2008zn},
SAGE \cite{Abdurashitov:2009tn}, SNO \cite{Aharmim:2008kc}
or MINOS \cite{Adamson:2010uj} can be found in Ref.~\cite{Schwetz:2011qt}.}. 

In the SM exist various ways to explain neutrino data, e.g.\
postulating very heavy particles as in the seesaw mechanism \cite{seesaw,Cheng:1980qt}
resulting in a unique dimension-5 operator
\cite{Weinberg:1979sa}
or via loop-effects \cite{Cheng:1980qt,Zee:1985id,Ma:1998dn}.
Although all mentioned solutions
can be supersymmetrized, there exists an entirely supersymmetric
possibility to generate Majorana neutrino masses, namely
$R$-parity violation \rpv \cite{Aulakh:1982yn,Hall:1983id,Ross:1984yg}.
Allowing for all possible gauge- and supersymmetric invariant
terms within the MSSM particle content, $R$-parity can be broken
by bilinear or trilinear terms \cite{Hall:1983id}.
Trilinear models provide a huge number of free parameters, where
different attempts tried to reduce
them \cite{Drees:1997id}.
Spontaneous $R$-parity violation can be understood as a violation
of lepton number by the vacuum expectation value of some singlet
field \cite{Aulakh:1982yn,Masiero:1990uj,Amsler:2008zz}.
Bilinear \rpv can be interpreted as the low-energy 
limit of some spontaneous \rpv model, where the new singlet fields are all 
decoupled.
The bilinear model has only six new \rpv parameters 
and is thus more predictive in particular in view of collider physics
once the parameters have been chosen to explain correctly neutrino
data.
For an early work on bilinear $R$-parity violation
in the context of a low-energy effective theory we refer to \cite{Lee:1984kr}.

The phenomenology of \rpv SUSY has been studied extensively in 
the past, for reviews see \cite{Barbier:2004ez,Chemtob:2004xr,Hirsch:2004he}. 
Neutrino masses have been calculated with trilinear couplings 
\cite{Hall:1983id,Dedes:2006ni} and for pure bilinear models 
\cite{Roy:1996bua,Hempfling,Bisset:1998bt,Hirsch:2000ef,Diaz:2003as} or
both \cite{Davidson:2000ne}.
Neutrino angles are 
not predicted in either schemes, but can be easily fitted to 
experimental data. In bilinear schemes the requirement to correctly 
explain neutrino data fixes all \rpv couplings in sufficiently small 
intervals, so that in some specific final states of the decays of 
the LSP correlations with neutrino angles appear. This has been 
shown for a (bino-dominated) neutralino LSP in
 \cite{Mukhopadhyaya:1998xj,Porod:2000hv,Chun:2002rh}, 
for charged scalar LSPs in \cite{Hirsch:2002ys}, for sneutrino
LSPs in \cite{Hirsch:2003fe,Aristizabal Sierra:2004cy}, and for 
chargino, gluino and squark LSPs in \cite{Hirsch:2003fe}. Such 
a tight connection between neutrino physics and LSP decays is 
lost to some extent in the general trilinear-plus-bilinear case.

The $\mu$-problem of the MSSM \cite{Kim:1983dt} is induced by
the mass term for the Higgs superfields $\mu \widehat{H}_d \widehat{H}_u$.
For phenomenological reasons the parameter $\mu$ must be of the order
of the electroweak scale. However, if there is a larger scale in the
theory, the natural value of $\mu$ should be around this large scale.
A solution to this problem can be provided by the Next-to-Minimal
SSM (NMSSM) \cite{Barbieri:1982eh,Nilles:1982dy}, where an additional
singlet superfield is introduced. The vacuum expectation value
of its scalar component
generates an effective $\mu$ term as soon as electroweak symmetry
is broken. The phenomenology has been discussed for example in
\cite{Djouadi:2008uw,Ellwanger:2009dp}.
Electroweak mass corrections were discussed in \cite{Staub:2010ty},
an on-shell scheme for neutralino and chargino masses including
the discussion of the two-body decays $\tilde{\chi}_i^0\rightarrow
\tilde{\chi}_j^\pm W^\mp$ at one-loop level was worked out in \cite{Liebler:2010bi}.

The $\mu\nu$SSM, which was first proposed in \cite{LopezFogliani:2005yw},
uses the same singlet superfield(s) not only to generate the $\mu$ term
but in addition Dirac mass terms for the left-handed neutrinos.
The cubic self-couplings of the singlets in the 
superpotential explicitly break 
lepton number and moreover also $R_p$. Thus, Majorana neutrino
masses are generated once electroweak symmetry is broken.
One-loop corrections to the neutrino mass matrix have
been calculated in \cite{Bartl:2009an,Ghosh:2010zi}.
Compared to the NMSSM the $\mu\nu$SSM with one right-handed
neutrino superfield also contains six new \rpv parameters, the same number
as going from the MSSM to the bilinear \rpv model as will be discussed
in detail later on. 
This implies that one also obtains correlations
between neutrino mixing angles and ratios of LSP decay
branching ratios \cite{Bartl:2009an,Ghosh:2008yh}.
Moreover, one has unusual decay modes and signatures in the Higgs 
sector \cite{Bartl:2009an,Bandyopadhyay:2010cu}.

Although one expects correlations between neutrino mixing angles and
final states of the LSP decays at tree level, it is a priori unclear 
what happens
at one-loop level. In particular using one-loop on-shell masses
and the one-loop mixing matrix together with the
tree level decay widths results in case of the $\mu\nu$SSM with
one right-handed neutrino superfield in an unexpected behavior:
The expected correlation does not necessarily show up for a singlino-like
LSP. Therefore, we present in this paper  a full one-loop
calculation for the decays $\tilde{\chi}_1^0\rightarrow l^\pm W^\mp$
and show how the tree level expectations evolve at full one-loop level.
For this aim we extend the on-shell scheme  presented in
\cite{Liebler:2010bi} for the  decays 
$\tilde{\chi}_i^0\rightarrow \tilde{\chi}_j^\pm W^\mp$ within
the MSSM and the NMSSM to the models considered here.
Moreover, we briefly discuss the implementation of these decays 
in \prog{} \cite{CNNDecaysonline}.
Note that in case of the $\mu\nu$SSM with
more than one right-handed neutrino superfield
similar problems do not show up. The reason is, that neutrino
physics can be fully explained at tree level and thus at
the same level of perturbation theory as the correlations
with the neutralino branching ratios.

This paper is organized as follows: We first define our models
in Section~\ref{sec:modeldefinition}.
In Section~\ref{sec:theoryoneloopmasses} we present our on-shell
scheme for the neutralino and chargino mass matrices in the considered
models. Afterwards we present numerical results for the neutrino
masses and mixings comparing the on-shell scheme and the $\overline{\text{DR}}$
scheme in 
Section~\ref{sec:practiceoneloopmasses}. 
Next we discuss the decays  $\tilde{\chi}_1^0\rightarrow l^\pm W^\mp$
and the correlations to neutrino physics at tree level
in Section~\ref{sec:treelevelwidth}. In Section~\ref{sec:oneloopwidth}
we first discuss the complete NLO corrections to the widths
and discuss the correlations at one-loop  level.
Finally we conclude in Section~\ref{sec:conclusion}.
\ref{sec:CNNDecaysapp} gives the implementation of the
considered models in the public program \prog.

\section{Model definition}
\label{sec:modeldefinition}

In this section we introduce the models by presenting the superpotential
and the soft-breaking terms. Moreover we will give the neutralino and
chargino mass matrices at tree level and comment on the need of
one-loop corrections for the neutrino masses.

\subsection{Superpotential, soft terms and scalar potential}
\label{subsec:superpot}

We will start with the superpotential of bilinear $R$-parity violation
(BRpV) and the $\mu\nu$SSM with
one right-handed neutrino superfield~$\widehat \nu^c$.
Both model have in common the Yukawa part of the
matter superfields coupled to the $SU(2)$-doublet
Higgs fields
{\allowdisplaybreaks\begin{align}
{\cal W}_{Y} =  \epsilon_{\alpha\beta} \left(Y_U^{ij}  \widehat Q_i^\alpha \widehat u_j^c \widehat H_u^\beta
+ Y_D^{ij} \widehat Q_i^\beta   \widehat d_j^c \widehat H_d^\alpha
+  Y_l^{ij} \widehat L_i^\beta \widehat l_j^c \widehat H_d^\alpha \right)\qquad,
\label{eq:WsuppotY}
\end{align}}\xspaceskip 0pt
where $\epsilon_{\alpha\beta}$ is the complete antisymmetric $SU(2)$ tensor with $\epsilon_{12}=1$
and $i,j$ denote the three families.
In case of BRpV the explicit $\mu$ and $\epsilon_i$-terms are added
and the total superpotential reads as
\vspace{-4mm}
{\allowdisplaybreaks\begin{align}
{\cal W}_{b\text{\rpv}} = {\cal W}_{Y} + \epsilon_{\alpha\beta}\left[\epsilon_i\widehat{L}_i^\alpha\widehat{H}_u^\beta
- \mu \widehat{H}_d^\alpha \widehat{H}_u^\beta \right]\qquad,
\label{eq:Wsuppotbrpv}
\end{align}}\xspaceskip 0pt
whereas in the $\mu\nu$SSM the Higgs doublets are coupled to the right-handed neutrino
superfield~$\widehat \nu^c$ and in addition a term containing the neutrino Yukawa coupling $Y_\nu$ is added
{\allowdisplaybreaks\begin{align}
{\cal W}_{\mu\nu\text{SSM}} = &{\cal W}_{Y}
 +  \epsilon_{\alpha\beta}\left[Y_{\nu}^{i} \widehat{L}_i^\alpha \widehat{\nu}^c \widehat{H}_u^\beta
 - \lambda \widehat{\nu}^c \widehat{H}_d^\alpha \widehat{H}_u^\beta\right]
 +\frac{1}{3}\kappa \widehat{\nu}^c \widehat{\nu}^c \widehat{\nu}^c\qquad.
\label{eq:WsuppotmunuSSM}
\end{align}}\xspaceskip 0pt
The first term in Equation~\eqref{eq:Wsuppotbrpv} and the last two terms
in Equation~\eqref{eq:WsuppotmunuSSM} explicitly
break lepton number if one assigns lepton number to $\widehat{\nu}^c$.
If not the first term in Equation~\eqref{eq:WsuppotmunuSSM} explicitly
breaks lepton number and one immediately sees that in both models
one has the same number of \rpv violating parameters. Note that
for the phenomenology it does not matter if $\widehat{\nu}^c$ carries
lepton number as it is broken explicitly by a least one interaction of
this field.
Both models have in common the soft SUSY breaking terms
{\allowdisplaybreaks\begin{align}\nonumber
 V_{soft}^{\text{MSSM} - B\mu} = &{m_Q^{ij}}^2 \tilde{Q}_i^{\alpha \ast}
\tilde{Q}_j^\alpha + {m_U^{ij}}^2 \tilde{u}_i^c \tilde{u}_j^{c\ast} +
{m_D^{ij}}^2 \tilde{d}_i^c \tilde{d}_j^{c\ast} + {m_L^{ij}}^2
\tilde{L}_i^{\alpha \ast} \tilde{L}_j^\alpha + {m_E^{ij}}^2 \tilde{l}_i^c
\tilde{l}_j^{c\ast} \\\nonumber & +  m_{H_d}^2 H_d^{a \ast} H_d^a +
m_{H_u}^2 H_u^{a \ast} H_u^a - \frac{1}{2} \big[ M_1 \tilde{B}^0
\tilde{B}^0 + M_2 \tilde{W}^\gamma \tilde{W}^\gamma + M_3 \tilde{g}^{\gamma'}
\tilde{g}^{\gamma'} + h.c. \big] \label{eq:softmssm} \\ &+  \epsilon_{\alpha\beta}
\big[ T_U^{ij} \tilde{Q}_i^\alpha \tilde{u}_j^{c\ast} H_u^\beta + T_D^{ij}
\tilde{Q}_i^\beta \tilde{d}_j^{c\ast} H_d^\alpha + T_E^{ij} \tilde{L}_i^\beta
\tilde{l}_j^{c\ast} H_d^\alpha + h.c. \big]\quad,
\end{align}}\xspaceskip 0pt
where $V_{soft}^{\text{MSSM} - B\mu}$ contains all the usual soft terms of the MSSM
but the $B_\mu$-term and a summation over $\gamma=1,\ldots,3$ and $\gamma'=1,\ldots,8$ has to be performed. 
In addition we add in case of BRpV
{\allowdisplaybreaks\begin{align}
V_{soft}^{b\text{\rpv}} = V_{soft}^{\text{MSSM} - B\mu} + \epsilon_{\alpha\beta} \left[
B_i\epsilon_i\tilde{L}_i^\alpha H_u^\beta
- B \mu H_d^\alpha H_u^\beta
+h.c.\right]
\end{align}}\xspaceskip 0pt
and in case of the $\mu\nu$SSM:
{\allowdisplaybreaks\begin{align}\nonumber
V_{soft}^{\mu\nu\text{SSM}} = &V_{soft}^{\text{MSSM} - B\mu} + {m_{\tilde{\nu}^c}}^2 \tilde{\nu}^c
  \tilde{\nu}^{c \ast} \\ &+ \epsilon_{\alpha\beta} \big[ T_{\nu}^i
  \tilde{L}_i^\alpha \tilde{\nu}^c H_u^\beta - T_\lambda \tilde{\nu}^c
  H_d^\alpha H_u^\beta + h.c. \big] + \big[\frac{1}{3}T_{\kappa} \tilde{\nu}^c \tilde{\nu}^c
  \tilde{\nu}^c +h.c.\big]
\end{align}}\xspaceskip 0pt
In these expressions the notation for the soft trilinear couplings
introduced in \cite{Skands:2003cj} is used.
We will not present the scalar potential and the corresponding
tadpole equations in this paper, but refer to \cite{Hirsch:2000ef} for BRpV
and to \cite{Bartl:2009an} for the $\mu\nu$SSM. Note the different
definition of $\kappa$ in comparison to \cite{Bartl:2009an}.
The determination of the vacuum structure through the scalar potential induces
the following vacuum expectation values (VEVs):
{\allowdisplaybreaks\begin{align}
\left< H_d^0\right>=\frac{1}{\sqrt{2}} v_d,\qquad
\left< H_u^0\right>=\frac{1}{\sqrt{2}} v_u,\qquad
\left< \tilde{\nu}_i\right>=\frac{1}{\sqrt{2}} v_i,\qquad
\left< \tilde{\nu}^c\right>=\frac{1}{\sqrt{2}} v_c
\end{align}}\xspaceskip 0pt
Again $\tilde{\nu}^c$ is only present in the $\mu\nu$SSM, where
its vacuum expectation value generates
effective $\mu$- and $\epsilon_i$-terms
{\allowdisplaybreaks\begin{align}
 \mu=\frac{1}{\sqrt{2}}\lambda v_c,\qquad
\epsilon_i=\frac{1}{\sqrt{2}}Y_\nu^i v_c\qquad.
\label{eq:muepseff}
\end{align}}\xspaceskip 0pt
In this way $\widehat \nu^c$ plays the role of the singlet superfield $\widehat S$ in
the NMSSM, implying that $\mu$, but also $\epsilon_i$ is naturally
of the order of the electroweak scale
\cite{Ellwanger:2009dp}. Moreover there are no additional terms with
dimension of mass present in the superpotential. Beside the parameter~$\epsilon_i$
we define the so-called alignment parameters
{\allowdisplaybreaks\begin{align}
\Lambda_i=\mu v_i+v_d\epsilon_i\qquad,
\label{eq:lambdadefinition}
\end{align}}\xspaceskip 0pt
which determine the neutrino mass matrix at tree level \cite{Bartl:2009an}
and which give the relevant information on the tree level mass
matrix in a basis invariant way \cite{Ferrandis:1998ii}.
$\epsilon_i$ and $\Lambda_i$ can then
be used to fit the neutrino data.
Concerning the scalar potential we add,
that in case of BRpV we determine the soft-breaking
parameters $m_{H_d}^2,m_{H_u}^2$ and $B_{i}$ and in
case of the $\mu\nu$SSM $m_{H_d}^2,m_{H_u}^2,m_{\nu^c}^2$ and
$T_\nu^i$ from the tadpole equations.

\subsection{Masses of neutralinos and charginos}
\label{subsec:neutralinoscharginos}

Using Weyl spinors defined by
{\allowdisplaybreaks\begin{align}
\left(\psi^-\right)^T=\left(\tilde{W}^-,\tilde{H}_d^-,e,\mu,\tau\right),\qquad
\left(\psi^+\right)^T=\left(\tilde{W}^+,\tilde{H}_u^+,e^c,\mu^c,\tau^c\right)
\end{align}}\xspaceskip 0pt
the Lagrangian density containing the chargino masses can be written in the form
{\allowdisplaybreaks\begin{align}
 \Lag\supset-\frac{1}{2}\left((\psi^-)^T\Mass_c \psi^+ + (\psi^+)^T(\Mass_c)^T\psi^-\right) + h.c.\qquad,
\end{align}}\xspaceskip 0pt
where the mass matrix of the charged fermions is given by
\begin{equation}
\Mass_c = \begin{pmatrix}
M_2 & \frac{1}{\sqrt{2}}gv_u & 0 & 0 & 0 \\
\frac{1}{\sqrt{2}}gv_d & \mu & -\frac{1}{\sqrt{2}}Y_l^{i1}v_i & -\frac{1}{\sqrt{2}}Y_l^{i2}v_i & -\frac{1}{\sqrt{2}}Y_l^{i3}v_i \\
\frac{1}{\sqrt{2}}gv_1 & -\epsilon_1 & \frac{1}{\sqrt{2}}Y_l^{11}v_d & \frac{1}{\sqrt{2}}Y_l^{12}v_d & \frac{1}{\sqrt{2}}Y_l^{13}v_d \\
\frac{1}{\sqrt{2}}gv_2 & -\epsilon_2 & \frac{1}{\sqrt{2}}Y_l^{21}v_d & \frac{1}{\sqrt{2}}Y_l^{22}v_d & \frac{1}{\sqrt{2}}Y_l^{23}v_d \\
\frac{1}{\sqrt{2}}gv_3 & -\epsilon_3 & \frac{1}{\sqrt{2}}Y_l^{31}v_d & \frac{1}{\sqrt{2}}Y_l^{32}v_d & \frac{1}{\sqrt{2}}Y_l^{33}v_d 
\end{pmatrix}\quad.
\end{equation}
The mass eigenstates are obtained by two unitary matrices $U$ and $V$ 
{\allowdisplaybreaks\begin{align}
F_i^+=V_{it}\psi_t^+ \quad\text{and}\quad F_i^-=U_{it}\psi_t^-\quad,
\end{align}}\xspaceskip 0pt
so that the mass eigenvalues can be calculated via
\begin{eqnarray}
\Mass_{c,dia.}=\text{Diag}\left(m_{\tilde{\chi}_1^+},\ldots,m_{\tilde{\chi}_5^+}\right)&=&U^*\mathcal{M}_c V^{-1}
\\
\nonumber
m_{\tilde{\chi}_i^+} \le m_{\tilde{\chi}_j^+}\qquad\text{for}\qquad i<j
\end{eqnarray}
and the Dirac spinor $\tilde{\chi}_i^+$ can be deduced from
\begin{equation}
\tilde{\chi}_i^+ = \left( \begin{array}{c}
F_i^+ \\ \overline{F^-_i}
\end{array} \right)\qquad.
\end{equation}
For the neutralinos we obtain in the basis \footnote{$\widehat \nu^c$ is only present
in the $\mu\nu$SSM.}
{\allowdisplaybreaks\begin{align}
\left( \psi^0 \right)^T &= \left( {\tilde B}^0, {\tilde W}_3^0,
 {\tilde H}_d^0, {\tilde H}_u^0, \nu^c, \nu_1,\nu_2,\nu_3 \right)
\label{eq:defbasis}
\end{align}}\xspaceskip 0pt
the mass matrices of the neutral fermions, which have the generic structure
{\allowdisplaybreaks\begin{align}
\Mass_n = \begin{pmatrix} M_H & \hat{m} \\ \hat{m}^T & 0\end{pmatrix}
\label{eq:neutralmass}
\end{align}}\xspaceskip 0pt
and enter the Lagrangian density $\Lag\supset -\frac{1}{2}\left((\psi^0)^T\Mass_n\psi^0\right)+h.c.$.
The submatrix $M_H$ contains the $4(5)$ heavy states, which are the four
usual neutralinos of the MSSM and in case of the $\mu\nu$SSM in addition
the right-handed neutrino. The submatrix $\hat{m}$ mixes the heavy states
with the left-handed neutrinos and contains the $R$-parity breaking
parameters. In detail the elements are given by
{\allowdisplaybreaks\begin{align}
M_H = \begin{pmatrix}
M_1 & 0 & -\frac{1}{2}g' v_d & \frac{1}{2}g' v_u & 0\\
0 & M_2 & \frac{1}{2}g v_d & -\frac{1}{2}g v_u & 0\\
-\frac{1}{2}g' v_d & \frac{1}{2}g v_d & 0 & -\mu & -\frac{1}{\sqrt{2}} \lambda v_u\\
\frac{1}{2}g' v_u & -\frac{1}{2}g v_u & -\mu & 0 & -\frac{1}{\sqrt{2}}\lambda v_d\\
0&0&-\frac{1}{\sqrt{2}} \lambda v_u&-\frac{1}{\sqrt{2}}\lambda v_d & \sqrt{2} \kappa v_c \end{pmatrix}\\
\hat{m}^T = \begin{pmatrix}
-\frac{1}{2} g' v_1 & \frac{1}{2} g v_1 & 0 & \epsilon_1 & \epsilon_1\frac{v_u}{v_c} \\
-\frac{1}{2} g' v_2 & \frac{1}{2} g v_2 & 0 & \epsilon_2 & \epsilon_2\frac{v_u}{v_c} \\
-\frac{1}{2} g' v_3 & \frac{1}{2} g v_3 & 0 & \epsilon_3 & \epsilon_3\frac{v_u}{v_c} \end{pmatrix}\qquad,
\label{eq:chi0masstree}
\end{align}}\xspaceskip 0pt
where in the $\mu\nu$SSM $\mu$ and $\epsilon_i$ can be taken from Equation~\eqref{eq:muepseff}.
In case of BRpV one has to omit the last column and the last row in
the submatrix $M_H$ and the last column in the matrix $\hat{m}$.
The mass eigenstates $F_i^0$ are obtained via
\begin{equation}
F_i^0=\mathcal{N}_{is}\psi_s^0 
\end{equation}
from the gauge eigenstates $\psi_s^0$, where the unitary matrix $\Nge$ diagonalizes 
the mass matrix $\Mass_n$ as
\begin{equation}
\Mass_{n,dia.}=\text{Diag}\left(m_{\tilde{\chi}_1^0},\ldots,m_{\tilde{\chi}_j^0}\right)
=\mathcal{N}^*\mathcal{M}_n\mathcal{N}^\dagger\qquad,
\end{equation}
where $j=7(8)$ in case of BRpV ($\mu\nu$SSM). The masses are ordered
as $m_{\tilde{\chi}_i^0} \le m_{\tilde{\chi}_j^0}$ for $i<j$ and we have
chosen $\Nge$ such that all mass eigenvalues are positive. This implies that
in general the matrix $\Nge$ is complex. Similar to \cite{Liebler:2010bi}
we can do useful checks using negative mass eigenvalues in combination with
a real mixing matrix $\Nge$.
The $4$-component spinor can be obtained via
\begin{equation}
\tilde{\chi}_i^0 = \left( \begin{array}{c}
F_i^0 \\ \overline{F^0_i}
\end{array} \right)\qquad.
\end{equation}
All the other masses of the scalar sectors can be taken from \cite{Hirsch:2000ef}
for BRpV and from \cite{Bartl:2009an} for the $\mu\nu$SSM.
The various couplings, which are needed for the one-loop calculations,
can be found in the folder couplings within \prog{} \cite{CNNDecaysonline}.

\subsection{Neutrino masses}
\label{subsec:neutrinomass}

Due to the smallness of the $R$-parity breaking parameters
an effective neutrino mass matrix in a seesaw approximation
can be found
\begin{equation}\label{eq:defmnueff}
m_{\nu\nu}^{\rm eff} = - \hat{m}^T M_H^{-1} \hat{m}
 = - \xi \hat{m}\quad,
\end{equation}
where the matrix $\xi$ contains the small expansion parameters and
can be taken from \cite{Hirsch:2000ef,Bartl:2009an}.
The basic calculation yields
{\allowdisplaybreaks\begin{align}\label{eq:effone}
(m_{\nu\nu}^{\rm eff})_{ij} &= a \Lambda_i \Lambda_j \qquad\text{with } a \text{ being}\\
a^{b\text{\rpv}}&=\frac{m_\gamma}{4\mathrm{Det}_0^{b\text{\rpv}}},\qquad
a^{\mu\nu\text{SSM}}=\frac{m_\gamma(\lambda^2v_dv_u+\mu m_R)}{4\mu \mathrm{Det}_0^{\mu\nu\text{SSM}}}\quad,
\end{align}}\xspaceskip 0pt
where we have used the abbreviations in Equations~\eqref{eq:muepseff} and \eqref{eq:lambdadefinition} and
{\allowdisplaybreaks\begin{align}
\label{eq:abbreviationsapproxcoupl}
m_\gamma = g^2M_1 +g'^2 M_2,\qquad \qquad m_R = \sqrt{2} \kappa v_c\quad,
\end{align}}\xspaceskip 0pt
where the singlet mass parameter $m_R$ is only valid in the $\mu\nu$SSM.
The determinants of $M_H$ are given by
{\allowdisplaybreaks\begin{align} \label{eq:det1}
\mathrm{Det}_0^{b\text{\rpv}} &= \frac{1}{2}m_\gamma\mu v_dv_u-M_1M_2\mu^2\\\label{eq:det2}
\mathrm{Det}_0^{\mu\nu\text{SSM}} &= \frac{1}{8} m_\gamma (\lambda^2 v^4 + 4 m_R \mu v_d v_u) 
- M_1 M_2 \mu(v_d v_u \lambda^2 + m_R \mu)\quad.
\end{align}}\xspaceskip 0pt
The projective form of this mass matrix 
implies that only one neutrino acquires a tree level mass, while the other
two remain massless. We denote the matrix, which diagonalizes $m_{\nu\nu}^{\rm eff}$,
in the following $\Nge_\nu$.
The correct magnitude of $\vec{\Lambda}$ can be easily estimated from the approximated formula
for the neutrino mass $m_{\nu_3}$ at tree level
{\allowdisplaybreaks\begin{align}
\label{eq:approxlambda}
m_{\nu_3} = a|\vec{\Lambda}|\qquad.
\end{align}}\xspaceskip 0pt
Since one neutrino mass is not sufficient to explain the
oscillation data, one-loop corrections in BRpV \cite{Hirsch:2000ef,Diaz:2003as,Romao:1999up}
and in the $\mu\nu$SSM \cite{Bartl:2009an,Ghosh:2010zi,Das:2010wp} have to be added to match
the current bounds. 
We will therefore consider the complete mass matrices
at the  one-loop level using an on-shell scheme for calculating
the masses. In addition we will show, that this also requires
 the complete NLO corrections to the
decays $\Gamma(\tilde{\chi}_1^0\rightarrow l^\pm W^\mp)$ 
to obtain the correct correlations between neutrino mixing
angles and ratios of branching ratios.
 
\section{One-loop on-shell masses of neutralinos and charginos}
\label{sec:theoryoneloopmasses}

As we have shown pure tree level neutralino masses including the neutrinos
cannot explain the full neutrino bounds in BRpV and the $\mu\nu$SSM
with one right-handed neutrino $\widehat{\nu}^c$ only.
We will therefore go to the next order in perturbation theory for the masses,
which was illustrated using a $\overline{\text{DR}}$ scheme in BRpV
\cite{Hirsch:2000ef,Diaz:2003as,Romao:1999up}
and in the $\mu\nu$SSM \cite{Bartl:2009an,Ghosh:2010zi,Das:2010wp} in previous articles.
However, we will later focus on the decay width $\Gamma(\tilde{\chi}_1^0\rightarrow l^\pm_i W^\mp)$
at one-loop level to investigate the relations between branching ratios and neutrino
mixing angles on higher loop order and choose a more convenient on-shell scheme for our
calculation. Before discussing this topic, we will first discuss
the masses for neutralinos and charginos in an on-shell scheme
as in \cite{Liebler:2010bi}, which allows to have finite
corrections to the neutrino masses $m_{\nu_i}^{1L}$ but also to the mixing matrix $\Nge^{1L}$ at
one-loop level, which is crucial to explain the neutrino oscillation data.
The approach is similar to \cite{Eberl:2001eu,Oller:2003ge}, whereas
for the MSSM another procedure has been worked out in \cite{Fritzsche:2002bi,Baro:2008bg}.

\subsection{Common basic principles}

A special feature of the chargino/neutralino sector is, that the number of
parameters is lower than the number of imposed on-shell conditions. This
implies finite corrections to the tree level masses of neutralinos and charginos.
We closely follow \cite{Liebler:2010bi,Eberl:2001eu} and start with the one-loop contributions to
neutralino masses
$\delta \Mass_n^\circledast$ and chargino masses $\delta\Mass_c^\circledast$
{\allowdisplaybreaks\begin{align}
\label{eq:onshellmasses1}
&\left(\delta \Mass_n^\circledast\right)_{ij}=\delta\left(\Nge^T\Mass_{n,dia.}^\circledast \Nge\right)_{ij}\\\nonumber
&\quad=\sum_{n,l}\left[\delta\Nge_{ni}\left(\Mass_{n,dia.}^\circledast\right)_{nl}\Nge_{lj}+\Nge_{ni}\left(\delta\Mass_{n,dia.}^\circledast\right)_{nl}\Nge_{lj}+\Nge_{ni}\left(\Mass_{n,dia.}^\circledast\right)_{nl}\delta\Nge_{lj}\right]\\
\label{eq:onshellmasses2}
&\left(\delta \Mass_c^\circledast\right)_{ij}=\delta\left(U^T\Mass_{c,dia.}^\circledast V\right)_{ij}\\\nonumber
&\quad=\sum_{n,l}\left[\delta U_{ni}\left(\Mass_{c,dia.}^\circledast\right)_{nl}V_{lj}+U_{ni}\left(\delta\Mass_{c,dia.}^\circledast\right)_{nl}V_{lj}+U_{ni}\left(\Mass_{c,dia.}^\circledast\right)_{nl}\delta V_{lj}\right]
\end{align}}\xspaceskip 0pt
with the diagonalized mass matrices $\Mass_{dia.}^\circledast$ and their counterterms $\delta \Mass_{dia.}^\circledast$:
{\allowdisplaybreaks\begin{align}
\left(\Mass_{n,dia.}^\circledast\right)_{nl}=\delta_{nl}m_{\tilde{\chi}_l^0}&,\quad \left(\delta \Mass_{n,dia.}^\circledast\right)_{nl}
=\delta_{nl}\delta m_{\tilde{\chi}_l^0}\\ \left(\Mass_{c,dia.}^\circledast\right)_{nl}=\delta_{nl}m_{\tilde{\chi}_l^\pm}&,
\quad \left(\delta \Mass_{c,dia.}^\circledast\right)_{nl}=\delta_{nl}\delta m_{\tilde{\chi}_l^\pm}
\end{align}}\xspaceskip 0pt
The bare mass matrices of the neutralinos and charginos can now be expressed as the corrected on-shell mass matrix
with the corrections in Equations~\eqref{eq:onshellmasses1} and \eqref{eq:onshellmasses2} or via the tree level mass matrix
expressed in physical parameters together with the renormalization constants of those:
{\allowdisplaybreaks\begin{align}
\Mass_{n,c}^0=\Mass_{n,c}^\circledast+\delta \Mass_{n,c}^\circledast=\Mass_{n,c}+\delta\Mass_{n,c}
\end{align}}\xspaceskip 0pt
Therefore the relations between tree level and one-loop mass matrices take the form:
{\allowdisplaybreaks\begin{align}
\label{eq:oneloopmasses}
\Mass_{n,c}^\circledast&=\Mass_{n,c}+\delta\Mass_{n,c}-\delta\Mass_{n,c}^\circledast=:\Mass_{n,c}+\varDelta \Mass_{n,c}
\end{align}}\xspaceskip 0pt
In the following we will define the model-dependent physical parameters, write down the
renormalization of the mass matrices $\delta\Mass_{n,c}$ and identify the renormalization constants
of the physical parameters,
which should be fixed in the neutralino or chargino sector.
Some physical parameters, namely $\delta m_W,
\delta m_Z$ and thus $\delta\cos\theta_W$ are fixed in the gauge boson sector
and $\delta\tan\beta$ in the Higgs sector. In particular for
$\delta\tan\beta$ we take
the $\overline{\text{DR}}$ renormalization~\cite{Freitas:2002um}, so that
UV divergences in the masses and the considered process cancel
{\allowdisplaybreaks\begin{align}
 \frac{\delta\tan\beta}{\tan\beta}=\frac{1}{32\pi^2}\Delta\left( 3\text{Tr}({Y_d}{Y_d}^\dagger)
 - 3\text{Tr}({Y_u}{Y_u}^\dagger) + \text{Tr}({Y_l}{Y_l}^\dagger) - \text{Tr}({Y_\nu}{Y_\nu}^\dagger) \right) \qquad,
\end{align}}\xspaceskip 0pt
where $\Delta=\tfrac{1}{\epsilon}-\gamma +\ln(4\pi)$ parametrizes the UV divergence as in \cite{Liebler:2010bi}
and $Y_\nu$ is only present in the $\mu\nu$SSM.
Note that this choice maintains also the gauge invariance of masses and
the considered decay widths. We will also need the renormalization constants
for the sine and cosine of different angles, which are given by:
{\allowdisplaybreaks\begin{align}
 \frac{\delta\cos\beta}{\cos\beta}=-\sin^2\beta\frac{\delta\tan\beta}{\tan\beta},\quad
 \frac{\delta\sin\beta}{\sin\beta}=\cos^2\beta\frac{\delta\tan\beta}{\tan\beta}
\end{align}}\xspaceskip 0pt

\subsection{Bilinear $R$-parity breaking}

The on-shell renormalization of neutralino and chargino masses in the MSSM and NMSSM
was extensively discussed in \cite{Liebler:2010bi}.
In case of BRpV the renormalization of the physical parameters is more challenging.
The masses of the $Z$ and $W$ bosons are defined in the following form
{\allowdisplaybreaks\begin{align}
 m_Z^2=\frac{1}{4}(g^2+g'^2)\left(v_d^2+v_u^2+\sum_i v_i^2\right),
\qquad m_W^2=\frac{1}{4}g^2 \left(v_d^2+v_u^2+\sum_i v_i^2\right)\quad,
\end{align}}\xspaceskip 0pt
where not only the vacuum expectation values of the Higgs enter, but also the ones of the left-handed sneutrinos.
As in the (N)MSSM the $\cos$ of the given Weinberg angle $\cos\theta_W$ can be derived
from $m_W=\cos\theta_W m_Z$. 
Beside the already known $\tan\beta$,
we define 
{\allowdisplaybreaks\begin{align}
\tan \beta_i=\frac{v_i}{v_d} \qquad\text{and}\qquad \epsilon_i
\end{align}}\xspaceskip 0pt
as additional parameters, which are present at tree level, such that
the tree level mass matrix of the neutralinos is given by
{\allowdisplaybreaks\begin{align}
\Mass_n&=\begin{pmatrix}M_H&\hat{m}\\\hat{m}^T&0\end{pmatrix}\\
 M_H&=\begin{pmatrix}M_1&0&-m_Z\sin\theta_W\cos\beta\Theta&m_Z\sin\theta_W\sin\beta\Theta\\
&M_2&m_Z\cos\theta_W\cos\beta\Theta&-m_Z\cos\theta_W\sin\beta\Theta\\
&&0&-\mu\\
&\text{sym.}&&0\end{pmatrix}\\
(\hat{m}^T)_i&=\begin{pmatrix}-m_Z\sin\theta_W\cos\beta\tan\beta_i\Theta& m_Z\cos\theta_W\cos\beta\tan\beta_i\Theta & 0 &\epsilon_i\end{pmatrix}\quad,
\end{align}}\xspaceskip 0pt
where $\Theta$ is a function of $\tan\beta$ and $\tan\beta_i$, namely:
{\allowdisplaybreaks\begin{align}
\label{eq:Deltabetabetai}
\Theta=\Theta(\beta,\beta_i)=\sqrt{\frac{1}{1+\cos^2\beta\sum_i\tan^2\beta_i}}
\quad\stackrel{v_i\rightarrow 0}{\longrightarrow}\quad 1
\end{align}}\xspaceskip 0pt
In this way we maintain the possibility to fix the renormalization constants of $m_Z$ and $\cos\theta_W$ in the
gauge boson sector, whereas the corrections from $R$-parity breaking are parameterized by $\tan\beta_i$,
$\epsilon_i$ and the parameter $\Theta$, the latter depends on $\beta$ and $\beta_i$.
Defining in addition the lepton masses $m_{ij}=\tfrac{1}{\sqrt{2}}Y_l^{ij}v_d$ the chargino mass matrix is given by:
{\allowdisplaybreaks\begin{align}
\label{eq:massmatrixcharginosbilinear}
\Mass_c=\begin{pmatrix}M_2&\sqrt{2}m_W\sin\beta\Theta&0&0&0\\
\sqrt{2}m_W\cos\beta\Theta&\mu&-\tan\beta_im_{i1}&-\tan\beta_im_{i2}&-\tan\beta_im_{i3}\\
\sqrt{2}m_W\cos\beta\tan\beta_1\Theta&-\epsilon_1&m_{11}&m_{12}&m_{13}\\
\sqrt{2}m_W\cos\beta\tan\beta_2\Theta&-\epsilon_2&m_{21}&m_{22}&m_{23}\\
\sqrt{2}m_W\cos\beta\tan\beta_3\Theta&-\epsilon_3&m_{31}&m_{32}&m_{33}\end{pmatrix}
\end{align}}\xspaceskip 0pt
We discuss the general case of non-diagonal lepton masses $m_{ij}$, since the
counterterms $\delta m_{ij}$ are needed to cancel non-diagonal contributions
in the lepton mass matrix at one-loop level. However, the lepton Yukawa
couplings $Y_l$ and thus $m_{ij}$ can be chosen diagonal at tree level.
Using now the relations $\cos\beta^0=\cos\beta+\delta\cos\beta$ and $\tan\beta_i^0=\tan\beta_i+\delta\tan\beta_i$
one can do a simple expansion in first order of $\delta\cos\beta$ and $\delta\tan\beta_i$:
{\allowdisplaybreaks\begin{align}\nonumber
\Theta_0&=\sqrt{\frac{1}{1+\cos^2\beta^0\sum_i\tan^2\beta_i^0}}\approx \sqrt{\frac{1}{1+\cos^2\beta\sum_i\tan^2\beta_i}}
\\\nonumber&\quad-\left(\frac{1}{1+\cos^2\beta\sum_j\tan^2\beta_j}\right)^{\tfrac{3}{2}}
\sum_i(\cos^2\beta\tan\beta_i\delta\tan\beta_i+\cos\beta\delta\cos\beta\tan^2\beta_i)
\\&=\Theta+\delta\Theta
\end{align}}\xspaceskip 0pt
Therefore the counterterm of $\Theta$ can be expressed as a function of $\delta \tan\beta_i$ and $\delta\cos\beta$:
{\allowdisplaybreaks\begin{align}
\delta \Theta&=-\cos\beta\delta\cos\beta\Theta^3\sum_i\tan^2\beta_i-\cos^2\beta\Theta^3\sum_i\tan\beta_i\delta \tan\beta_i
 \\\nonumber&=-\sum_i\cos^2\beta\tan^2\beta_i\Theta^3\left(\frac{\delta\cos\beta}{\cos\beta}+\frac{\delta\tan\beta_i}{\tan\beta_i}\right)
\quad\stackrel{v_i\rightarrow 0}{\longrightarrow}\quad 0
\end{align}}\xspaceskip 0pt
The variation of all non-zero 
entries of the tree level mass matrix yields:
{\allowdisplaybreaks\begin{align}
\label{eq:phfixneutralinosFbilinear}
\delta \Mass_n^{11}&=\delta M_1=\frac{\delta M_1}{M_1}\Mass_n^{11}\\
\delta \Mass_n^{13}&=-\delta(m_Z\sin\theta_W\cos\beta\Theta)=\left(\frac{\delta m_Z}{m_Z}+\frac{\delta \sin\theta_W}{\sin\theta_W}+\frac{\delta \cos\beta}{\cos\beta}+\frac{\delta\Theta}{\Theta}\right)\Mass_n^{13}\\
\delta \Mass_n^{14}&=\delta(m_Z\sin\theta_W\sin\beta\Theta)=\left(\frac{\delta m_Z}{m_Z}+\frac{\delta \sin\theta_W}{\sin\theta_W}+\frac{\delta \sin\beta}{\sin\beta}+\frac{\delta\Theta}{\Theta}\right)\Mass_n^{14}\\
\delta \Mass_n^{1,4+i}&=-\delta(m_Z\sin\theta_W\cos\beta\tan\beta_i\Theta)\\\nonumber&=\left(\frac{\delta m_Z}{m_Z}+\frac{\delta \sin\theta_W}{\sin\theta_W}+\frac{\delta \cos\beta}{\cos\beta}+\frac{\delta\tan\beta_i}{\tan\beta_i}+\frac{\delta\Theta}{\Theta}\right)\Mass_n^{1,4+i}\\
\delta \Mass_n^{22}&=\delta M_2=\frac{\delta M_2}{M_2}\Mass_n^{22}\\
\delta \Mass_n^{23}&=\delta(m_Z\cos\theta_W\cos\beta\Theta)=\left(\frac{\delta m_Z}{m_Z}+\frac{\delta \cos\theta_W}{\cos\theta_W}+\frac{\delta \cos\beta}{\cos\beta}+\frac{\delta\Theta}{\Theta}\right)\Mass_n^{23}\\
\delta \Mass_n^{24}&=-\delta(m_Z\cos\theta_W\sin\beta\Theta)=\left(\frac{\delta m_Z}{m_Z}+\frac{\delta \cos\theta_W}{\cos\theta_W}+\frac{\delta \sin\beta}{\sin\beta}+\frac{\delta\Theta}{\Theta}\right)\Mass_n^{24}\\
\delta \Mass_n^{2,4+i}&=\delta(m_Z\cos\theta_W\cos\beta\tan\beta_i\Theta)\\\nonumber&=\left(\frac{\delta m_Z}{m_Z}+\frac{\delta \cos\theta_W}{\cos\theta_W}+\frac{\delta \cos\beta}{\cos\beta}+\frac{\delta\tan\beta_i}{\tan\beta_i}+\frac{\delta\Theta}{\Theta}\right)\Mass_n^{2,4+i}\\
\delta \Mass_n^{34}&=-\delta\mu=\frac{\delta\mu}{\mu}\Mass_n^{34}\\
\delta \Mass_n^{4,4+i}&=\delta\epsilon_i=\frac{\delta\epsilon_i}{\epsilon_i}\Mass_n^{44+i}
\label{eq:phfixneutralinosLbilinear}
\end{align}}\xspaceskip 0pt
whereas all the other variations $\delta\Mass_n^{12}=\delta\Mass_n^{33}=\delta\Mass_n^{3,4+i}=\delta\Mass_n^{44}=0$ necessarily vanish. The corrections in the chargino mass matrix read:
{\allowdisplaybreaks\begin{align}
\label{eq:phfixcharginosFbilinear}
\delta \Mass_c^{11}&=\delta M_2=\frac{\delta M_2}{M_2}\Mass_n^{22}\\
\delta \Mass_c^{12}&=\sqrt{2}\delta(m_W\sin\beta\Theta)=\left(\frac{\delta m_W}{m_W}+\frac{\delta \sin\beta}{\sin\beta}+\frac{\delta\Theta}{\Theta}\right)\Mass_c^{12}\\
\delta \Mass_c^{21}&=\sqrt{2}\delta(m_W\cos\beta\Theta)=\left(\frac{\delta m_W}{m_W}+\frac{\delta \cos\beta}{\cos\beta}+\frac{\delta\Theta}{\Theta}\right)\Mass_c^{21}\\
\delta \Mass_c^{22}&=\delta \mu=\frac{\delta \mu}{\mu}\Mass_c^{22}\\
\delta \Mass_c^{2,2+i}&=\delta\left(-\tan\beta_k m_{ki}\right)=\sum_k\left(\frac{\delta \tan\beta_k}{\tan\beta_k}-\frac{\delta m_{ki}}{m_{ki}}\right)\Mass_c^{2,2+i}\\
\delta \Mass_c^{2+i,1}&=\sqrt{2}\delta(m_W\cos\beta\tan\beta_i\Theta)\\\nonumber&=\left(\frac{\delta m_W}{m_W}+\frac{\delta \cos\beta}{\cos\beta}+\frac{\delta\tan\beta_i}{\tan\beta_i}+\frac{\delta\Theta}{\Theta}\right)\Mass_c^{2+i,1}\\
\delta \Mass_c^{2+i,2}&=-\delta\epsilon_i=\frac{\delta\epsilon_i}{\epsilon_i}\Mass_c^{2+i,2}\\
\delta \Mass_c^{2+i,2+j}&=\delta m_{ij}=\frac{\delta m_{ij}}{m_{ij}}\Mass_c^{2+i,2+j}
\label{eq:phfixcharginosLbilinear}
\end{align}}\xspaceskip 0pt
and $\delta\Mass_c^{1,2+i}=0$ vanishes. 
Similar to the (N)MSSM we fix $\delta M_1$ in the neutralino and $\delta M_2,\delta \mu$ in the chargino sector.
However, we still have to find appropriate renormalization conditions for $\delta \tan\beta_i,\delta \epsilon_i$
and $\delta m_{ij}$. Whereas $\delta m_{ij}$
should be fixed in the lepton sector, so that one-loop contributions to leptonic two-point functions
vanish, we have several possibilities for $\delta \tan\beta_i$ and $\delta \epsilon_i$. Whereas $\delta \tan\beta_i$
could for example be fixed in the Higgs sector together with  
$\tan\beta$,
one could fix $\delta\epsilon_i$ with respect to an $R$-parity violating decay.
We will focus on stable lepton masses. Thus, we calculate $\delta\tan\beta_i$ and $\delta\epsilon_i$ from
the one-loop contributions in the chargino sector, which mix the well-known MSSM charginos with the leptons.
In total we therefore impose the following conditions for the MSSM parameters
{\allowdisplaybreaks\begin{align}
\varDelta\Mass_n^{11}=\varDelta\Mass_c^{11}=\varDelta\Mass_c^{22}\stackrel{!}{=}0\qquad,
\end{align}}\xspaceskip 0pt
which results in
{\allowdisplaybreaks\begin{align}
\delta M_1&=\delta \Mass_n^{\circledast 11},\qquad \delta M_2=\delta \Mass_c^{\circledast 11},
\qquad \delta \mu=-\delta \Mass_c^{\circledast 22}\quad.
\end{align}}\xspaceskip 0pt
In a second step the conditions $\varDelta \Mass_c^{2+i,2+j}\stackrel{!}{=}0$ for the renormalization
constants of the lepton masses $\delta m_{ij}$ are imposed, resulting in
{\allowdisplaybreaks\begin{align}
\delta m_{ii}=\delta \Mass_c^{\circledast 2+i,2+j}\quad.
\end{align}}\xspaceskip 0pt
In a last step the $R$-parity violating sector with $\delta\tan\beta_i$ and $\delta\epsilon_i$
is considered, which can be fixed by imposing $\varDelta \Mass_c^{2+i,2}=\varDelta\Mass_c^{2,2+i}\stackrel{!}{=}0$
{\allowdisplaybreaks\begin{align}
&\delta\tan\beta_i=\frac{1}{\text{det}(m)}\frac{1}{2}\sum_{j,k,l,r,s}\epsilon_{ijk}\epsilon_{lrs}
\Upsilon_lm_{jr}m_{ks}\\
&\qquad\text{with}\quad \Upsilon_i=-\sum_k \tan\beta_k \delta m_{ki}-\delta\Mass_n^{\circledast 2,2+i}\\&\delta\epsilon_i=\delta\Mass_c^{\circledast 2+i,2}\quad,
\end{align}}\xspaceskip 0pt
where $\epsilon_{ijk}$ is the Levi-Civita symbol and det$(m)$ is the determinant of the lepton masses.
In case of vanishing non-diagonal lepton masses at tree level $m_{ij}=0$ for $i\neq j$ this simplifies
to $\delta\tan\beta_i =\tfrac{1}{m_{ii}}\Upsilon_i$.
Another possibility is to fix $\delta\tan\beta_i$ from $\varDelta \Mass_c^{2+i,1}$. However, this induces
a dependence on the renormalization constants of the gauge sector, whereas
in the described case the neutrino and lepton sector are ``decoupled'' from those. 

For all the remaining entries of the neutralino and chargino mass matrices shifts $\varDelta \Mass_{n,c}$
have to be taken into account. Due to the non-vanishing entries $\varDelta \Mass_c^{2+i,1}$ also
the lepton masses differ between tree and one-loop level. However, we will show that this difference
is tiny, for reasonable neutrino masses far below the experimental uncertainties. Note
that the Yukawa couplings of the leptons at tree level of course have to be adopted, so that the
tree level lepton masses fit the experimental known values.

\subsection{$\mu\nu$SSM}

Combining the already known procedures from the NMSSM and BRpV we define
{\allowdisplaybreaks\begin{align}
&\tan \beta_i=\frac{v_i}{v_d},\qquad \tan \beta_c=\frac{v_c}{v_u}\\
\mu=\frac{1}{\sqrt{2}}\lambda &v_c,\qquad m_R=\frac{1}{\sqrt{2}}\kappa v_c,\qquad \epsilon_i=\frac{1}{\sqrt{2}}Y^i_\nu v_c\quad.
\end{align}}\xspaceskip 0pt
In this notation the tree level mass matrix of the neutralinos is given by
{\allowdisplaybreaks\begin{align}
\Mass_n&=\begin{pmatrix}M_H&\hat{m}\\\hat{m}^T&0\end{pmatrix}\\
 M_H&=\begin{pmatrix}M_1&0&-m_Z\sin\theta_W\cos\beta\Theta&m_Z\sin\theta_W\sin\beta\Theta&0\\&M_2&m_Z\cos\theta_W\cos\beta\Theta&-m_Z\cos\theta_W\sin\beta\Theta&0\\&&0&-\mu&-\frac{\mu}{\tan\beta_c}\\&\text{sym.}&&0&\frac{\tan\beta_i\epsilon_i-\mu}{\tan\beta\tan\beta_c}\\&&&&m_R\end{pmatrix}\\
(\hat{m}^T)_i&=\begin{pmatrix}-m_Z\sin\theta_W\cos\beta\tan\beta_i\Theta& m_Z\cos\theta_W\cos\beta\tan\beta_i\Theta & 0 &\epsilon_i&\frac{\epsilon_i}{\tan\beta_c}\end{pmatrix}\quad,
\end{align}}\xspaceskip 0pt
where $\Theta$ is defined in Equation~\eqref{eq:Deltabetabetai} and the
chargino mass matrix has the same form as in BRpV,
see Equation~\eqref{eq:massmatrixcharginosbilinear}.
Beside the variations already present in the bilinear model as shown
in Equations~\eqref{eq:phfixneutralinosFbilinear}-\eqref{eq:phfixneutralinosLbilinear}, where the indices $4+i$ have to be shifted
to $5+i$, $\delta \Mass_n$ has the following additional entries:
{\allowdisplaybreaks\begin{align}
&\delta \Mass_n^{35}=\delta\left(-\frac{\mu}{\tan\beta_c}\right)=\left(\frac{\delta \mu}{\mu}-\frac{\delta \tan\beta_c}{\tan\beta_c}\right)\Mass_n^{35}\\
&\delta\Mass_n^{45}=\delta\left(\frac{\tan\beta_i\epsilon_i-\mu}{\tan\beta\tan\beta_c}\right)
=\left(-\frac{\delta\tan\beta}{\tan\beta}-\frac{\delta\tan\beta_c}{\tan\beta_c}+\frac{\delta\tan\beta_i\epsilon_i+\tan\beta_i\delta\epsilon_i-\delta\mu}{\tan\beta_i\epsilon_i-\mu}\right)\Mass_n^{45}\\
&\delta\Mass_n^{55}=\frac{\delta m_R}{m_R}\Mass_n^{55}\\
&\delta\Mass_n^{5,5+i}=\delta\left(\frac{\epsilon_i}{\tan\beta_c}\right)=\left(\frac{\delta\epsilon_i}{\epsilon_i}-\frac{\delta\tan\beta_c}{\tan\beta_c}\right)\Mass_n^{55+i}
\end{align}}\xspaceskip 0pt
whereas all the other variations $\delta\Mass_n^{12}=\delta\Mass_n^{15}=\delta\Mass_n^{25}=\delta\Mass_n^{33}=\delta\Mass_n^{3,5+i}=\delta\Mass_n^{44}=0$
necessarily vanish. The corrections in the chargino mass matrix are the same
as in the bilinear case, presented in Equations~\eqref{eq:phfixcharginosFbilinear}-\eqref{eq:phfixcharginosLbilinear}.
Similar to the MSSM, we fix $\delta M_1$ in the neutralino and $\delta M_2,\delta \mu$ in the chargino sector.
Similar to the NMSSM, we fix $\delta \tan\beta_c$ and $\delta m_R$ in the neutralino sector and similar
to the BRpV case $\delta\tan\beta_i$ and $\delta\epsilon_i$ are determined in
the chargino sector. However, we will summarize these results once again for the $\mu\nu$SSM.
We start with the conditions for the non-$R$-parity breaking variables
{\allowdisplaybreaks\begin{align}
\varDelta\Mass_c^{11}=\varDelta\Mass_c^{22}=\varDelta\Mass_n^{11}=\varDelta\Mass_n^{35}=\varDelta\Mass_n^{55}\stackrel{!}{=}0\quad,
\end{align}}\xspaceskip 0pt
which results in:
{\allowdisplaybreaks\begin{align}
\delta M_1&=\delta \Mass_n^{\circledast 11},\qquad \delta M_2=\delta \Mass_c^{\circledast 11},\qquad \delta \mu=\delta \Mass_c^{\circledast 22}\\
\delta \tan\beta_c &=\frac{\tan^2\beta_c}{\mu}\left(\delta \Mass_n^{\circledast 35}-\frac{1}{\tan\beta_c} \delta \Mass_n^{\circledast 34}\right),
\qquad \delta m_R=\delta \Mass_n^{\circledast 55}
\end{align}}\xspaceskip 0pt
In a second step the conditions
$\varDelta \Mass_c^{2+i,2}=\varDelta\Mass_c^{2,2+i}=\varDelta \Mass_c^{2+i,2+j}\stackrel{!}{=}0$
for the renormalization constants of the lepton masses $\delta m_{ij}$,
$\delta\tan\beta_i$ and $\delta\epsilon_i$ are imposed, resulting in:
{\allowdisplaybreaks\begin{align}
&\delta m_{ij}=\delta \Mass_c^{\circledast 2+i,2+j},\qquad \delta\epsilon_i=\delta\Mass_c^{\circledast 2+i,2}\\
&\delta\tan\beta_i=\frac{1}{\text{det}(m)}\frac{1}{2}\sum_{j,k,l,r,s}\epsilon_{ijk}\epsilon_{lrs}
\Upsilon_lm_{jr}m_{ks}\\
&\qquad\text{with}\quad \Upsilon_i=-\sum_k \tan\beta_k \delta m_{ki}-\delta\Mass_n^{\circledast 2,2+i}
&
\end{align}}\xspaceskip 0pt

\subsection{Definition of one-loop on-shell masses}
With the procedure introduced in the last sections we can calculate one-loop on-shell neutralino and chargino masses
for BRpV and the $\mu\nu$SSM, namely by combining the full one-loop corrections $\delta\Mass_{n,c}^\circledast$ 
with the counterterms $\delta \Mass_{n,c}$ obtained according to Equation~\eqref{eq:oneloopmasses}.
This results in the one-loop mass matrix $\Mass_{n,c}^\circledast$, whose diagonalizations lead to one-loop
neutralino $m^{1L}_{\tilde{\chi}_i^0}$ and chargino masses $m^{1L}_{\tilde{\chi}_i^\pm}$
and mixing matrices at the one-loop level $\Nge^{1L},U^{1L},V^{1L}$.
Note that these masses are UV and IR finite as well as gauge independent, if one takes into
account the gauge independent
renormalization of the mixing matrices as presented in \cite{Liebler:2010bi}
for the Equations~\eqref{eq:onshellmasses1} and \eqref{eq:onshellmasses2}.
Similar to the case of $\overline{\text{DR}}$ masses
an effective neutrino mass matrix of the form
{\allowdisplaybreaks\begin{align}
 (m_{\nu\nu}^{\rm eff})_{ij} &= a \Lambda_i \Lambda_j + b (\epsilon_i\Lambda_j + \epsilon_j\Lambda_i) + c \epsilon_i\epsilon_j
\label{eq:effneutrinomatrixoneloop}
\end{align}}\xspaceskip 0pt
is obtained at one-loop level \cite{Hirsch:2000ef,Diaz:2003as}.

\subsection{Calculation of the neutrino parameters}
With the one-loop on-shell masses $m^{1L}_{\nu_i}$ of the neutrinos and the mixing matrix $\Nge^{1L}$
one can calculate the relevant parameters to be compared with experimental neutrino data:
{\allowdisplaybreaks\begin{align}
\label{eq:massdifferences}
\Delta m^2_{atm} = \left(m^{1L}_{\nu_3}\right)^2 - \left(m^{1L}_{\nu_1}\right)^2,\qquad
\Delta m^2_{sol} = \left(m^{1L}_{\nu_2}\right)^2 - \left(m^{1L}_{\nu_1}\right)^2\\
\tan^2\theta_{atm} = \left|\frac{\Nge^{1L}_{3,6}}{\Nge^{1L}_{3,7}}\right|^2,\qquad
\tan^2\theta_{sol} = \left|\frac{\Nge^{1L}_{2,5}}{\Nge^{1L}_{1,5}}\right|^2,\qquad
U_{e3}^2 = \left|\Nge^{1L}_{3,5}\right|^2
\label{eq:mixingangles}
\end{align}}\xspaceskip 0pt
The formulas are valid for BRpV, in case of the $\mu\nu$SSM one has to
replace $\Nge^{1L}_{i,j}\rightarrow \Nge^{1L}_{i,j+1}$.

\section{One-loop on-shell neutrino masses}
\label{sec:practiceoneloopmasses}

In this section we show the results for the one-loop on-shell neutrino masses
in BRpV and the $\mu\nu$SSM, which we obtain using the definitions of the last sections.
The two models and the procedure explained before are also added to \prog{} \cite{CNNDecaysonline},
so that the reader is able to reproduce all the presented results.

\subsection{Parameter choice}

In the choice of our parameters we closely follow the strategy presented in
\cite{Bartl:2009an}. In Table~\ref{tab:neutrinobounds} we show the current bounds on neutrino data
taken from \cite{Schwetz:2011qt}, which will be used for our later analysis.
Moreover we note for the specific models:
\begin{itemize}
\item BRpV: We use the low-energy parameter sets of the well-known 
``Snowmass Points and Slopes'' (SPS) \cite{AguilarSaavedra:2005pw,Allanach:2002nj}
as benchmark scenarios. In the discussion of the decay widths we
refer to two particular scenarios, which are defined
in Table~\ref{tab:4scmunuSSM}. SPS~$2$ was changed such, that
a higgsino-like lightest neutralino is present by setting
$M_1=M_2=600$~GeV.
The additional parameters $\epsilon_i, v_i$ or
respectively $\epsilon_i,\Lambda_i$ are fixed by the neutrino
constraints, which can be found in Table~\ref{tab:neutrinobounds}, if
not mentioned to be chosen otherwise. The corresponding $B_i$ are
determined from the tadpole equations.
\item $\mu\nu$SSM: We refer to the low-energy parameter sets of the
NMSSM benchmark scenarios \cite{Djouadi:2008uw,Ellwanger:2008py}
named mSUGRA~$i$ or GMSB~$j$ based on the soft SUSY breaking mechanism.
Details can also be found within \cite{Liebler:2010bi}.
We add for completeness, that the VEV $v_c$ is fixed from $\mu$ and $\lambda$
using Equation~\eqref{eq:muepseff}.
In the discussion of the decay widths we use four scenarios,
which are defined in Table~\ref{tab:4scmunuSSM}
and which are chosen such, that the lightest neutralino is either bino, wino,
higgsino or singlino.
Concerning perturbativity and charge or color breaking minima we refer to
\cite{Bartl:2009an,Escudero:2008jg} and note that our choices of $\lambda,\kappa$
and $T_\lambda, T_\kappa$ respect the given constraints.
The additional parameters $v_i,Y_\nu^i$ or respectively $\epsilon_i,\Lambda_i$ are
fixed by the neutrino
constraints, which can be found in Table~\ref{tab:neutrinobounds}, if
not mentioned to be chosen otherwise. The corresponding $T_\nu^i$ are
determined from the tadpole equations.
\end{itemize}

\begin{table}[]
\begin{center}
\begin{tabular}{| p{4cm} || c | c |}
\hline
Parameter & Best fit & $2\sigma$  \\
\hline \hline
$\Delta m_{21}^2[10^{-5}\text{~eV}^2]$ & $7.59^{+0.20}_{-0.18}$ & $7.24-7.99$\\
$|\Delta m_{31}^2|[10^{-3}\text{~eV}^2]$ & $2.45^{+0.09}_{-0.09}$ & $2.28-2.64$ \\
$\tan^2\theta_{12}=\tan^2\theta_{sol}$ & $0.453^{+0.037}_{-0.031}$ & $0.39-0.54$ \\
$\tan^2\theta_{23}=\tan^2\theta_{atm}$ & $1.04^{+0.28}_{-0.22}$ & $0.69-1.56$\\
$\tan^2\theta_{13}=\tan^2\theta_{R}$ & $0.010^{+0.009}_{-0.006}$ & $\leq 0.028$\\
\hline
\end{tabular}
\end{center}
\caption{Current bounds on neutrino data taken from \cite{Schwetz:2011qt}.}
\label{tab:neutrinobounds}
\end{table}

\begin{table}[]
\begin{center}
\begin{tabular}{| p{0.4cm} p{0.7cm} || c | c || c | c | c | c |}
\hline
\multicolumn{2}{|c||}{} & SPS~$2'$ & SPS~$3$ & mSUGRA~$1$ & mSUGRA~$3'$ & mSUGRA~$4'$ & GMSB~$5$  \\ \hline\hline
Changes & & $\begin{matrix}M_1=M_2\\=600\text{~GeV}\end{matrix}$ & None & None &
 $M_1\leftrightarrow M_2$ & $\begin{matrix}\kappa \approx 0.1\\\rightarrow 0.4\end{matrix}$ & None\\ \hline\hline
$\tilde{\chi}^0_1:$ & $m$ & $371.05$ & $165.03$ & $210.79$ & $ 208.78$ & $196.82$ & $203.30$\\
 & $m^{1L}$               & $370.68$ & $164.76$ & $210.61$ & $208.55$ & $199.51$ & $203.30$\\
 & $C$                    & $\tilde{H}$ & $\tilde{B}$ & $\tilde{B}$ & $\tilde{W}$ & $\tilde{H}$&  $\tilde{S}$\\
 \hline
$\tilde{\chi}^0_2:$ & $m$ & $395.76$ & $295.36$ & $387.18$ & $391.11$ & $205.60$ &  $496.87$\\ 
 & $m^{1L}$               & $395.67$ & $295.73$ & $387.10$ & $390.82$ & $205.24$& $496.81$\\
 & $C$                    & $\tilde{H}$ & $\tilde{W}$ & $\tilde{W}$ & $\tilde{B}$ & $\tilde{H}$&  $\tilde{B}$\\
  \hline
$\tilde{\chi}^0_3:$ & $m$ & $600.00$ & $504.42$ & $971.11$ & $941.92$ & $327.26$ & $899.60$\\ 
 & $m^{1L}$               & $599.87$ & $504.76$ & $971.75$ & $940.00$ & $326.83$&  $899.98$\\ 
 & $C$                    & $\tilde{B}$ & $\tilde{H}$ & $\tilde{H}$ & $\tilde{H}$ & $\tilde{S}$&  $\tilde{W}$\\ 
 \hline
$\tilde{\chi}^0_4:$ & $m$ & $624.71$ & $521.86$ & $976.52$ & $942.49$ & $330.44$ &  $1377.67$\\ 
 & $m^{1L}$               & $624.42$ & $520.49$ & $975.14$ & $943.05$ & $330.95$&  $1377.45$\\ 
 & $C$                    & $\tilde{W}$ & $\tilde{H}$ & $\tilde{H}$ & $\tilde{H}$ & $\tilde{B}$&  $\tilde{H}$\\ 
 \hline
$\tilde{\chi}^0_5:$ & $m$ & & & $2101.57$ & $1421.70$ & $608.43$ &  $1383.97$\\ 
 & $m^{1L}$               & & & $2101.57$ & $1421.66$ & $607.65$&  $1383.04$\\ 
 & $C$                    & & & $\tilde{S}$ & $\tilde{S}$ & $\tilde{W}$&  $\tilde{H}$\\ 
  \hline
$\tilde{\chi}^\pm_1:$ & $m$ & $377.77$& $295.07$ & $387.16$ & $208.83$ & $201.36$&  $899.59$\\ 
 & $m^{1L}$                 & $378.22$& $295.58$ & $387.23$ & $208.77$ & $201.75$&  $900.14$\\ 
 & $C$                      & $\tilde{H}^\pm$ & $\tilde{W}^\pm$ & $\tilde{W}^\pm$ & $\tilde{W}^\pm$ & $\tilde{H}^\pm$&  $\tilde{W}^\pm$\\ 
  \hline
$\tilde{\chi}^\pm_2:$ & $m$ & $619.84$ & $521.90$ & $977.07$ & $946.04$ & $608.40$ &  $1384.24$\\ 
 & $m^{1L}$                 & $619.13$ & $521.03$ & $976.69$ & $945.71$ & $607.81$&  $1383.51$\\ 
 & $C$                      & $\tilde{W}^\pm$ & $\tilde{H}^\pm$ & $\tilde{H}^\pm$ & $\tilde{H}^\pm$ & $\tilde{W}^\pm$&  $\tilde{H}^\pm$\\ 
  \hline
\end{tabular}
\end{center}
\caption{$2$ scenarios for BRpV and $4$ scenarios for the $\mu\nu$SSM to allow for lightest neutralinos
with different particle characters $C$ where $m$ and $m^{1L}$ represent
the tree level and one-loop masses; SPS~$2$, mSUGRA~$3$ and $4$ were adopted,
so that also a wino- and higgsino-like lightest neutralino is accessible.}
\label{tab:4scmunuSSM}
\end{table}

\subsection{Absolute neutrino masses and mass differences}

First we illustrate the behavior of the absolute neutrino masses and
point out, that the on-shell renormalization allows a similar parameter
dependence as the $\overline{\text{DR}}$ masses, which
were defined in \cite{Hirsch:2000ef}.

\begin{figure}[ht]
\includegraphics[width=0.5\textwidth]{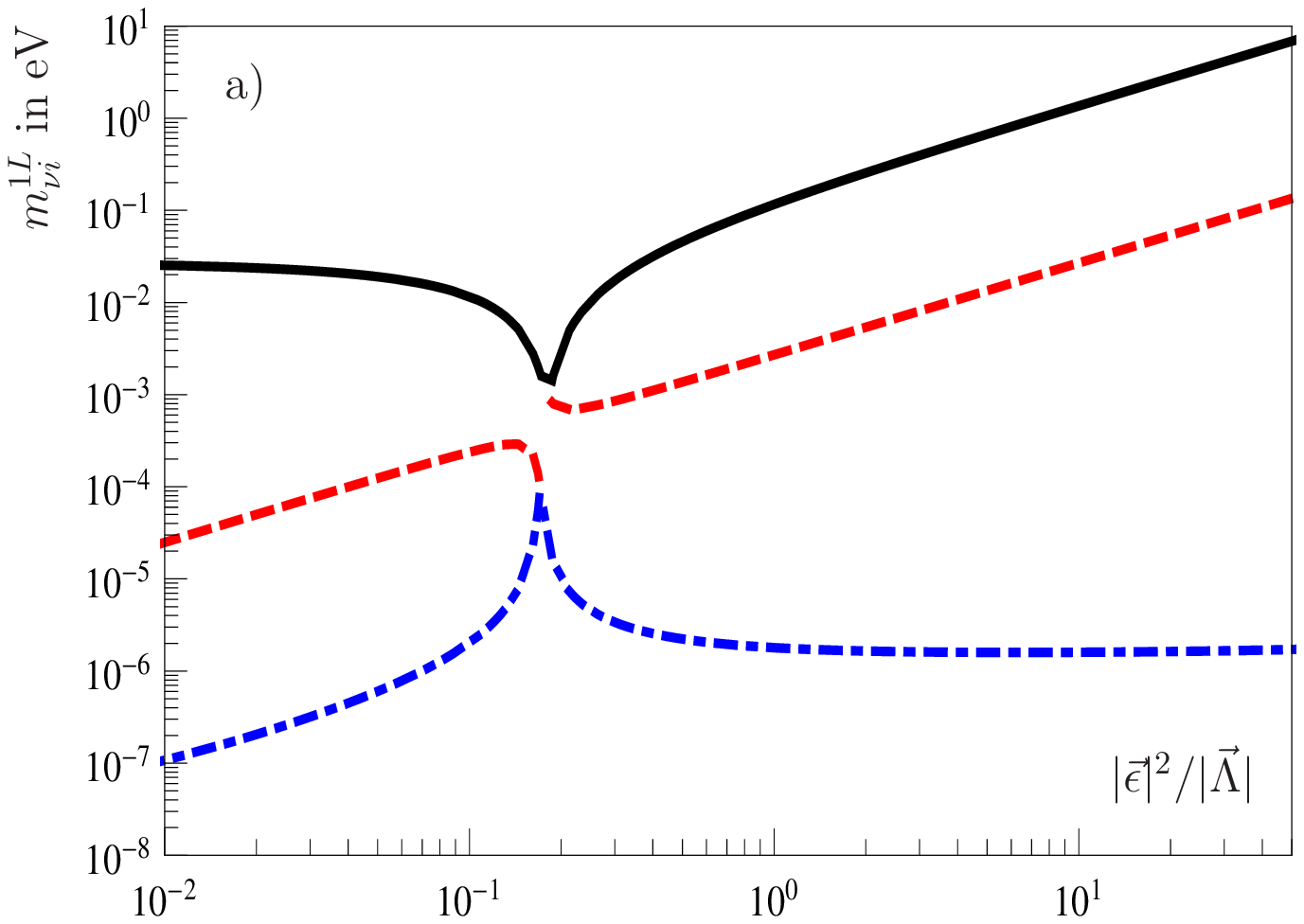}
\hfill
\includegraphics[width=0.5\textwidth]{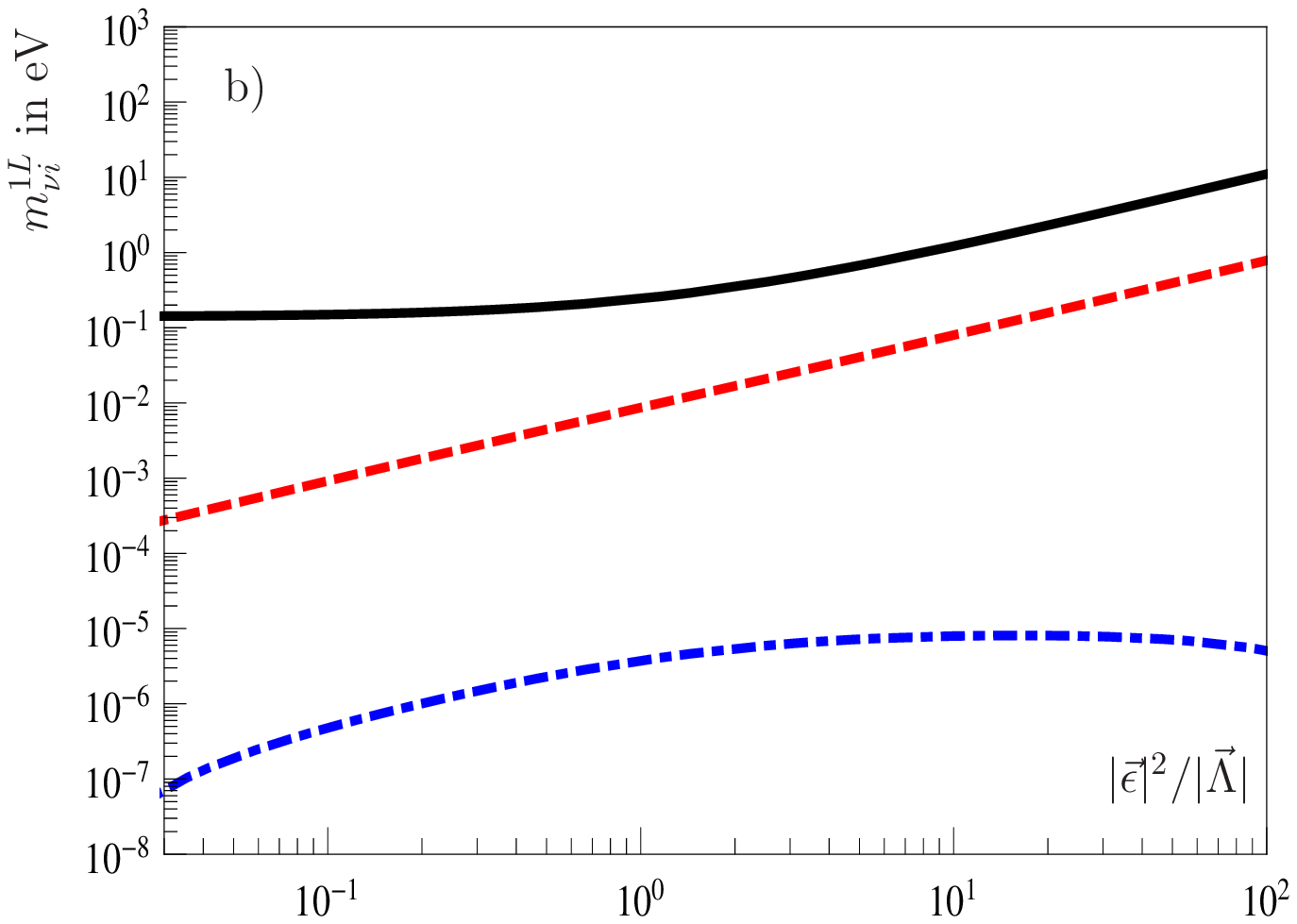}
\caption{
Three on-shell neutrino masses $m_{\nu_i}^{1L}$ in eV as a function of $|\vec{\epsilon}|^2/|\vec{\Lambda}|$
for a scenario in
a) the $\mu\nu$SSM based on mSUGRA~$1$;
b) BRpV based on SPS~$3$.
}
\label{fig:neumassabsolute}
\includegraphics[width=0.5\textwidth]{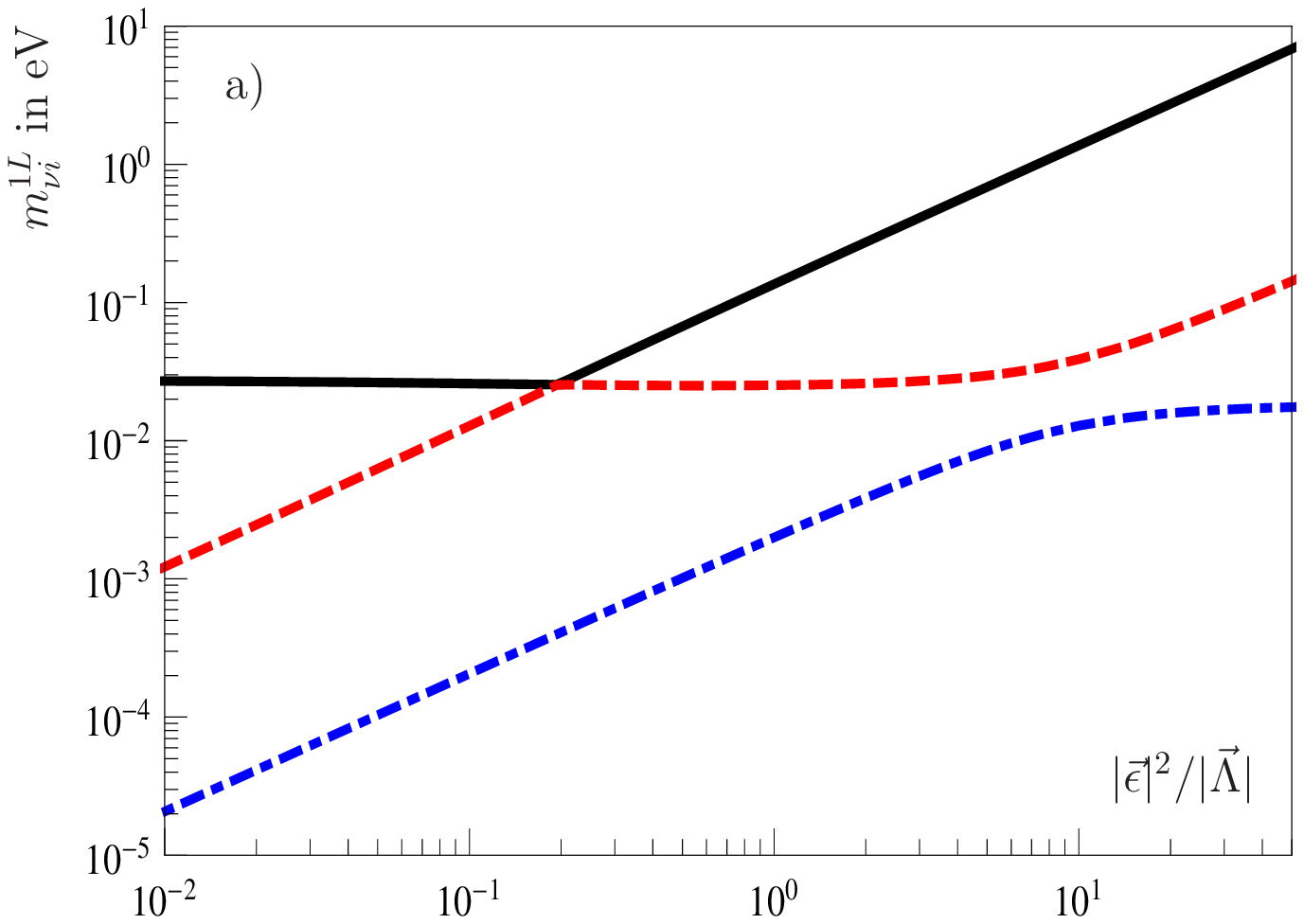}
\hfill
\includegraphics[width=0.5\textwidth]{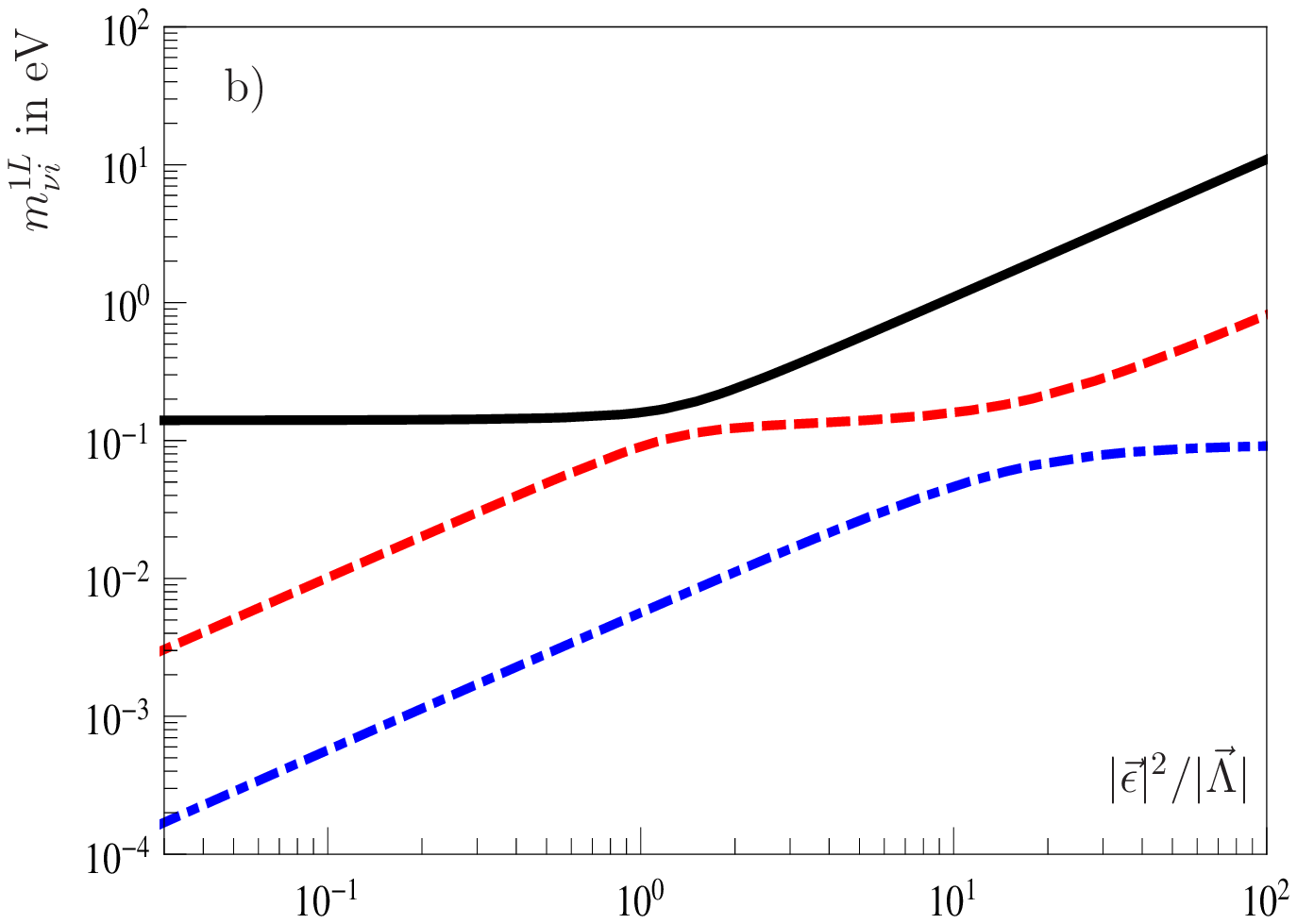}
\caption{
Three on-shell neutrino masses $m_{\nu_i}^{1L}$ in eV as a function of $|\vec{\epsilon}|^2/|\vec{\Lambda}|$
using the sign-condition defined in Equation~\eqref{eq:signcondition} for a scenario in
a) $\mu\nu$SSM based on mSUGRA~$1$;
b) BRpV based on SPS~$3$.
}
\label{fig:neumassabsolutesign}
\end{figure}

Figure~\ref{fig:neumassabsolute} shows the neutrino masses $m_{\nu_i}^{1L}$ of the three
left-handed neutrinos as a function of $|\vec{\epsilon}|^2/|\vec{\Lambda}|$, where
we have chosen $\epsilon_1=\epsilon_2=\epsilon_3$ and $\Lambda_1=\Lambda_2=\Lambda_3$
and a fixed value of $|\vec{\Lambda}|=0.235\text{~GeV}^2$. In case of the $\mu\nu$SSM
the scenario is based on mSUGRA~$1$ and in case of BRpV on SPS~$3$.
In Figure~\ref{fig:neumassabsolutesign} we set $\Lambda_1=-\Lambda_2=\Lambda_3$, so that
the sign-condition
{\allowdisplaybreaks\begin{align}
 \frac{\epsilon_2}{\epsilon_3}\frac{\Lambda_2}{\Lambda_3}<0
\label{eq:signcondition}
\end{align}}\xspaceskip 0pt
is fulfilled.
This allows for a simpler fit to the solar angle, since it helps to decouple the atmospheric
and the solar problem by reducing the contributions from the $b$-term
in the effective neutrino mass matrix at one-loop level in Equation~\eqref{eq:effneutrinomatrixoneloop} \cite{Hirsch:2000ef,Diaz:2003as}.
We will therefore make use of the sign-condition in the following.
Both models show a similar behavior.

The absolute value of $|\vec{\epsilon}|$ determines the neutrino masses $m_{\nu_1}^{1L}$
and $m_{\nu_2}^{1L}$, which are generated at one-loop level, whereas the tree level
neutrino mass $m_{\nu_3}^{1L}$ is set constant by $|\vec{\Lambda}|$
for small $|\vec{\epsilon}|$ and is only affected by the one-loop corrections
for large values of $|\vec{\epsilon}|$. Note that for the explanation
of the neutrino data the individual $\epsilon_i$ and $\Lambda_i$ have
to be chosen differently.
In case of the scenarios without sign-condition a level-crossing 
as in Figure~\ref{fig:neumassabsolute}~a)
can take place, which existence is dependent on the parameter point in the $\mu\nu$SSM
as well as in BRpV. It corresponds to a sign-flip between $m_{\nu_1}^{1L}$ and $m_{\nu_3}^{1L}$.

\begin{figure}[ht]
\includegraphics[width=0.5\textwidth]{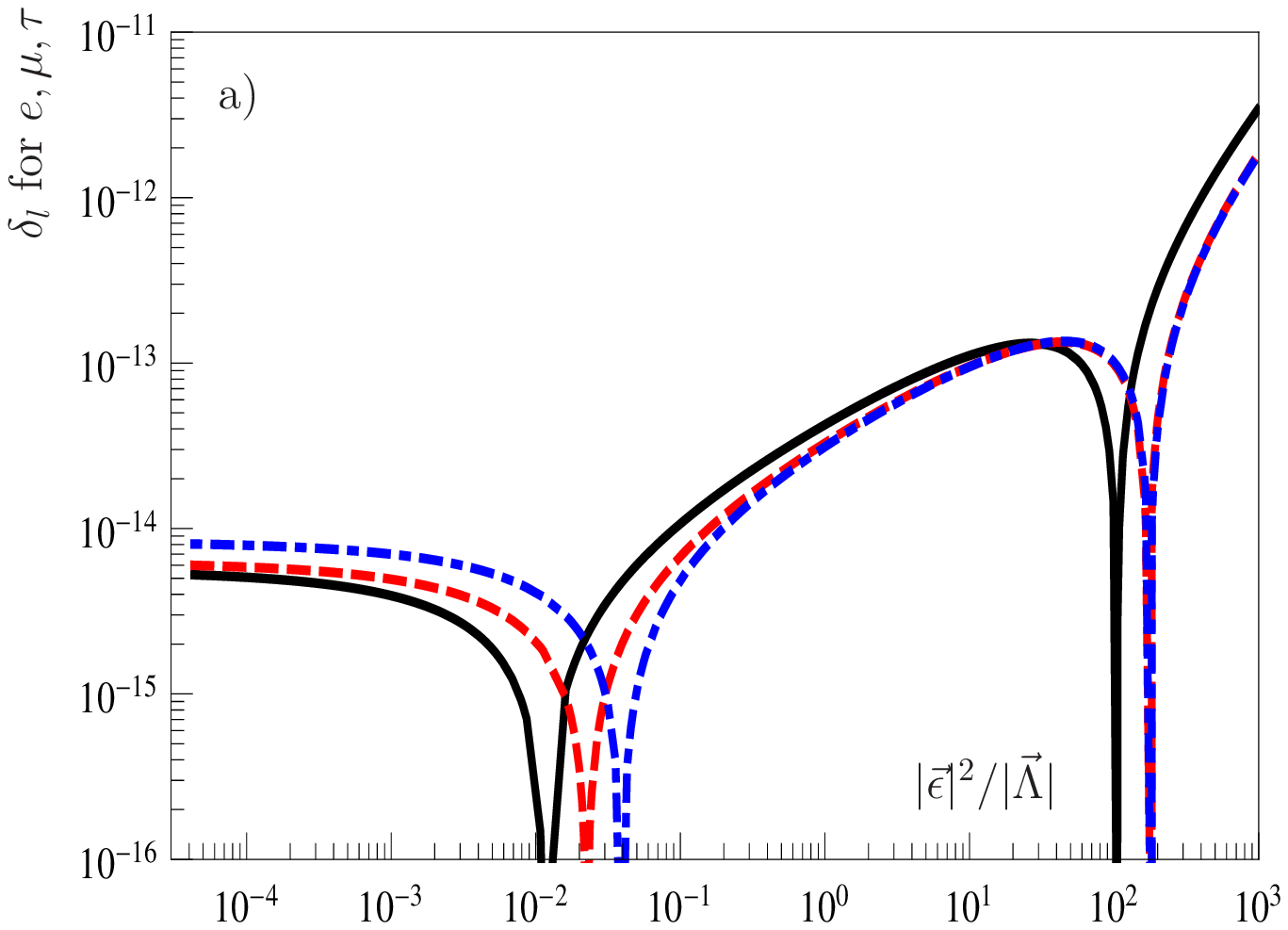}
\hfill
\includegraphics[width=0.5\textwidth]{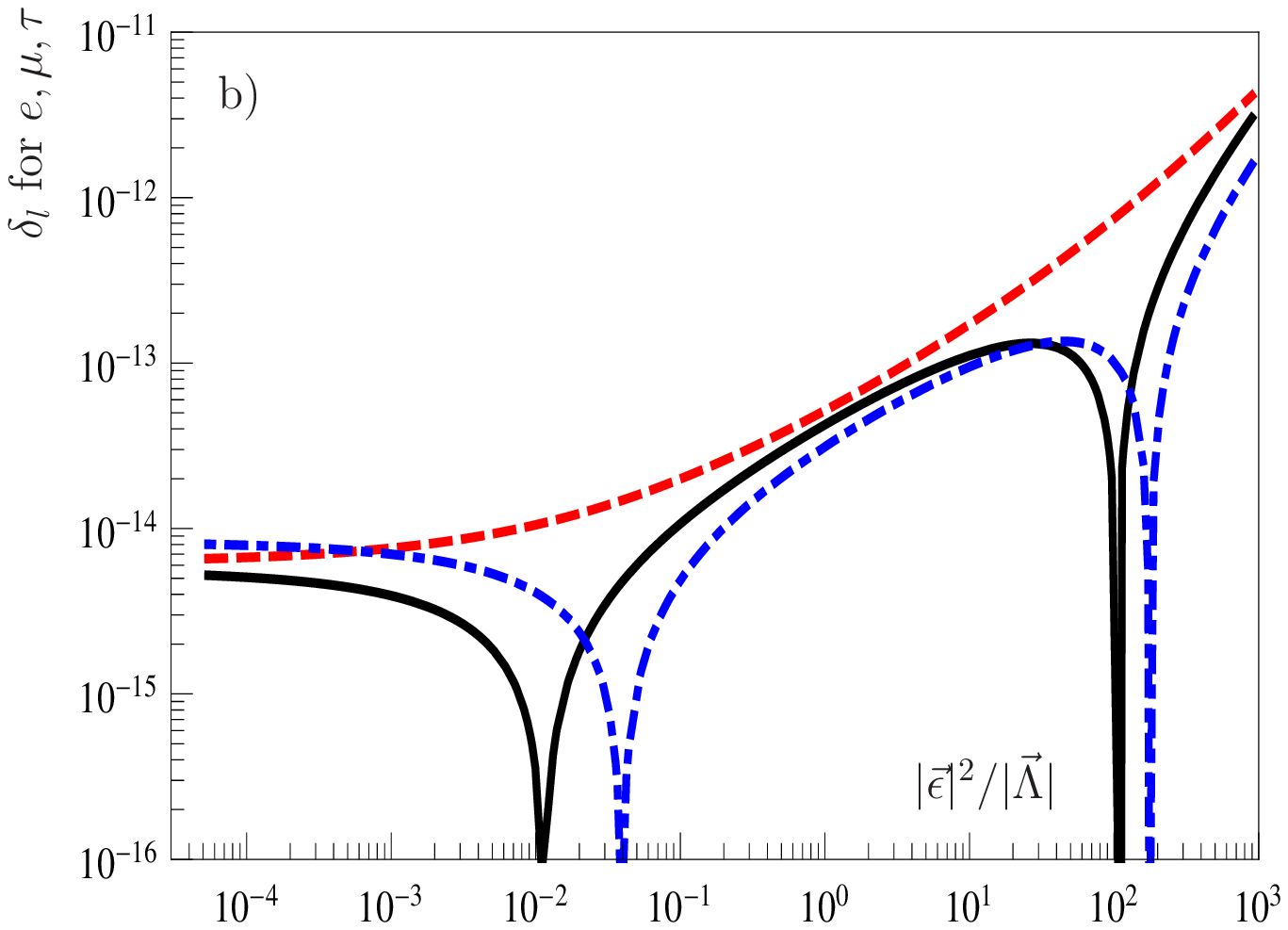}
\caption{
Corrections $\delta_l$ defined in Equation~\eqref{eq:deltalepmasses} for the three
lepton masses ($\tau$ (black, solid), $\mu$ (red, dashed), $e$ (blue,
dot-dashed)) as a function of $|\vec{\epsilon}|^2/|\vec{\Lambda}|$
for the $\mu\nu$SSM as in Figure~\ref{fig:neumassabsolute}~a), in detail
a) without sign-condition~\eqref{eq:signcondition};
b) with sign-condition~\eqref{eq:signcondition}.
}
\label{fig:lepmassabsolute}
\end{figure}

The Yukawa couplings $Y_l$ of the leptons at tree level
have to be adopted such, that the tree level lepton masses coincide with
the experimental values. Suppose we have fitted the tree level lepton masses $m_l$
to the experimental values, we can define the relative one-loop corrections by
{\allowdisplaybreaks\begin{align}
 \delta_l=\left|\frac{m^{1L}_l-m_l}{m_l}\right|\quad.
\label{eq:deltalepmasses}
\end{align}}\xspaceskip 0pt
For the $\mu\nu$SSM scenarios already presented in
Figures~\ref{fig:neumassabsolute}~a) and \ref{fig:neumassabsolutesign}~a) with
respect to the neutrino masses, Figure~\ref{fig:lepmassabsolute} shows
the relative correction $\delta_l$ for the lepton masses at one-loop level.
For reasonable neutrino masses the corrections to the
lepton masses are small $\delta_l<10^{-10}$,
even for the electron below the experimental uncertainties.
Note that the shown dips correspond to a sign change, since
we show the absolute value of the correction.

\begin{figure}[ht]
\includegraphics[width=0.5\textwidth]{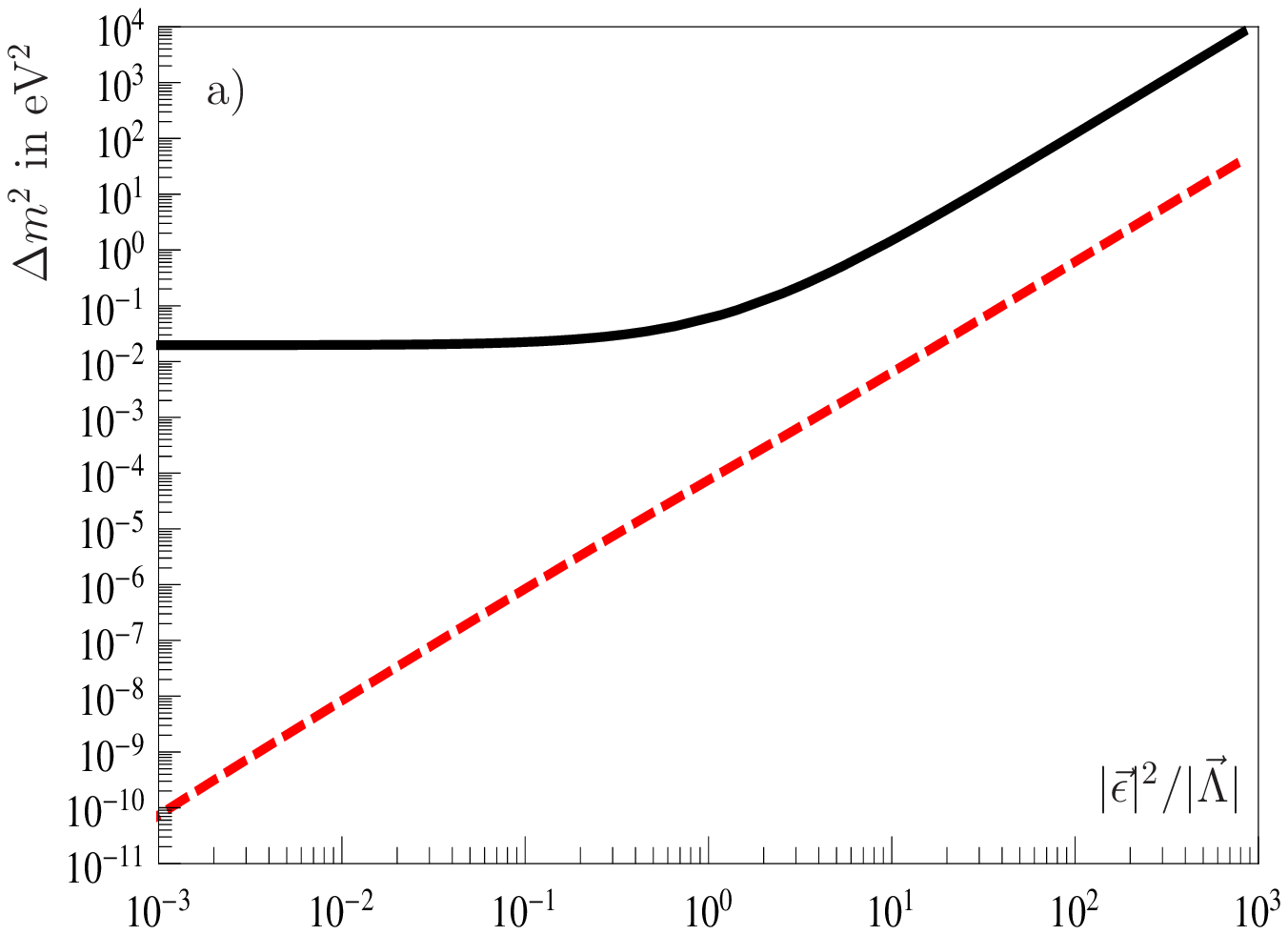}
\hfill
\includegraphics[width=0.5\textwidth]{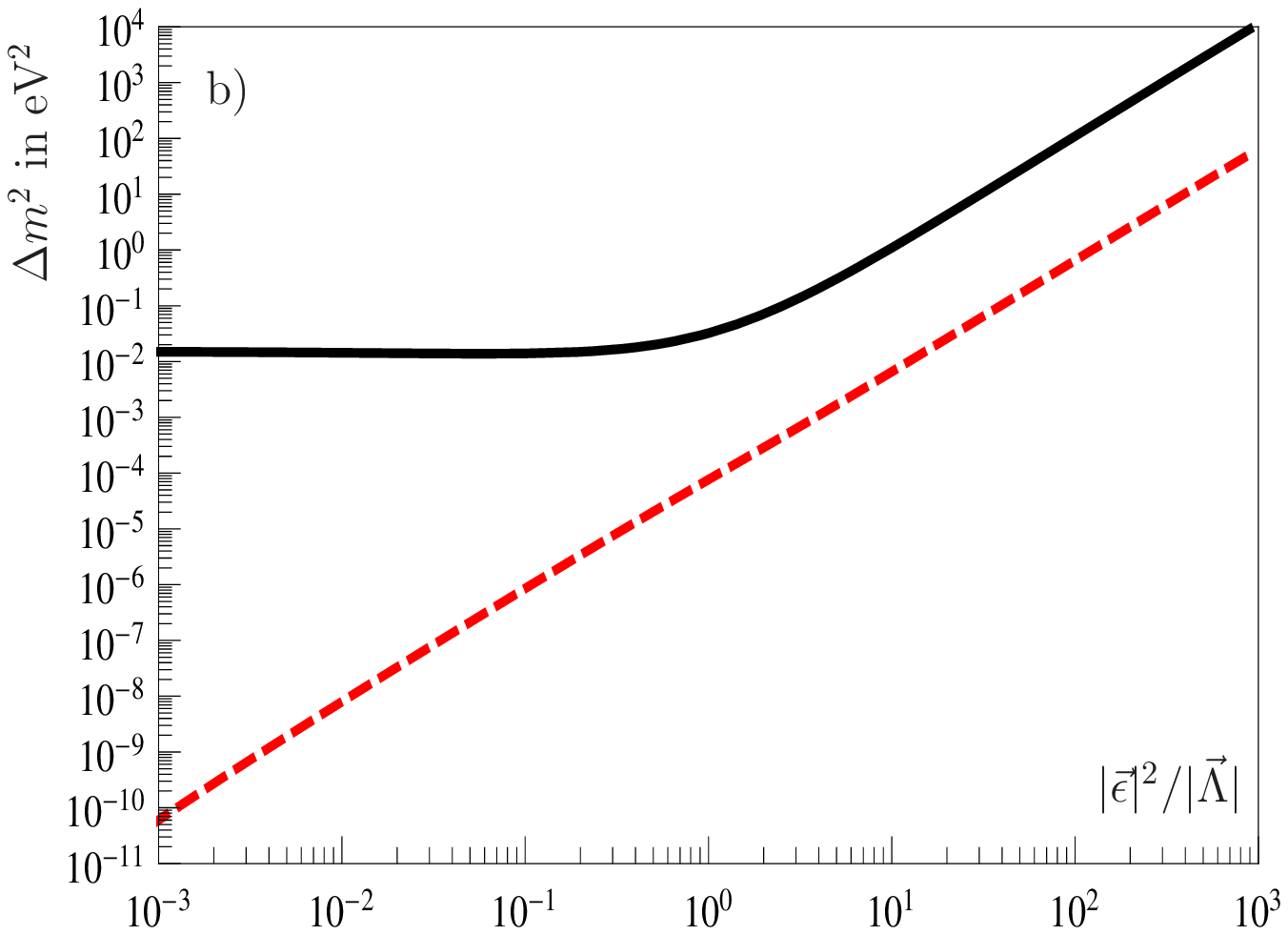}
\caption{
Mass differences $\Delta m^2_{atm}$ (black, solid) and $\Delta m^2_{sol}$ (red, dashed)
as a function of $|\vec{\epsilon}|^2/|\vec{\Lambda}|$ for BRpV
 based on SPS~$3$ using the
a) on-shell masses defined in Section \ref{sec:theoryoneloopmasses};
b) $\overline{\text{DR}}$ masses as defined in \cite{Hirsch:2000ef}.
}
\label{fig:BRPVm2MSOS}
\end{figure}

In the following we want to point out the difference between the
$\overline{\text{DR}}$ masses in \cite{Hirsch:2000ef} and the on-shell
masses as defined in this article. For BRpV
based on SPS~$3$ we present both schemes in Figure~\ref{fig:BRPVm2MSOS}.

Although there exists a difference
in the lightest neutrino mass $m_{\nu_1}^{1L}$, the mass differences
$\Delta m^2_{atm}$ and $\Delta m^2_{sol}$ defined in Equation~\eqref{eq:massdifferences}
are very similar, since those are given by
the absolute values of $m_{\nu_3}^{1L}$ and $m_{\nu_2}^{1L}$.
Since the on-shell procedure tends to reduce the impact
of the one-loop contributions, the largest mass $m_{\nu_3}^{1L}$ is in some
scenarios only for larger values of $|\vec{\epsilon}|$
affected by the one-loop contributions when comparing the on-shell with
the $\overline{\text{DR}}$ masses. However, for the neutrinos the two renormalization
prescriptions are very similar, whereas for the other neutralinos and charginos
and in particular for the leptons the corrections are (much) smaller in
case of on-shell masses compared to the $\overline{\text{DR}}$ masses.
In the scenarios above $\delta_l$ for $\overline{\text{DR}}$ lepton masses is between
$0.3$ and $2$\%. The one-loop corrections to the neutralino and chargino masses
are included in Table~\ref{tab:4scmunuSSM} and are typically in the
per-mil range, whereas in case of $\overline{\text{DR}}$ masses the corrections
are in the order of a few per-cent \cite{Staub:2010ty}.

\subsection{Relation between $\vec{\Lambda}$, $\vec{\epsilon}$ and the neutrino mass differences/the mixing angles}
\label{sec:rellambdaepsangles}

In this section we discuss the relations between the neutrino mass differences, the mixing angles
and the alignment parameters $\vec{\epsilon}$ and $\vec{\Lambda}$. We refer to \cite{Schwetz:2011qt}
for the current neutrino data, which we presented in Table~\ref{tab:neutrinobounds}.
For all figures presented in this section we fit $\vec{\epsilon}$ and $\vec{\Lambda}$ such,
that the neutrino data bounds are fulfilled except for the
angle shown in the figure.

Similar to \cite{Hirsch:2000ef} we use $\vec{\Lambda}$
at tree level to explain the atmospheric mass difference and the atmospheric mixing angle,
whereas $\vec{\epsilon}$ respectively $\vec{\tilde{\epsilon}}$ at one-loop level
is used for the solar mass difference and the solar mixing angle.
The latter correlation to $\vec{\tilde{\epsilon}}$ is more distinct than
the one to $\vec{\epsilon}$, where
a prerotation with the matrix $\Nge_\nu$ diagonalizing the
tree level neutrino mass matrix $m_{\nu\nu}^{\rm eff}$ given in Equation~\eqref{eq:effone}
according to $\vec{\tilde{\epsilon}}=\Nge_v \vec{\epsilon}$ was performed.
The mixing matrix $\Nge_\nu$ is in the considered scenarios exactly given by \cite{Hirsch:1998kc}
{\allowdisplaybreaks\begin{align}
 \Nge_\nu = \begin{pmatrix}\frac{\sqrt{\Lambda_2^2+\Lambda_3^2}}{|\vec{\Lambda}|} 
 & -\frac{\Lambda_1\Lambda_2}{\sqrt{\Lambda_2^2+\Lambda_3^2}|\vec{\Lambda}|}
 & -\frac{\Lambda_1\Lambda_3}{\sqrt{\Lambda_2^2+\Lambda_3^2}|\vec{\Lambda}|}\\
 0 & \frac{\Lambda_3}{\sqrt{\Lambda_2^2+\Lambda_3^2}}
 & -\frac{\Lambda_2}{\sqrt{\Lambda_2^2+\Lambda_3^2}} \\
 \frac{\Lambda_1}{|\vec{\Lambda}|} & \frac{\Lambda_2}{|\vec{\Lambda}|}
 & \frac{\Lambda_3}{|\vec{\Lambda}|}
\end{pmatrix}\quad .
\end{align}}\xspaceskip 0pt
We will first comment on the atmospheric and solar mass
differences $\Delta m^2_{atm}$ and $\Delta m^2_{sol}$.
Whereas the atmospheric mass difference is determined by $|\vec{\Lambda}|$,
the solar mass difference correlates with $|\vec{\epsilon}|$. Therein
the absolute value of $\vec{\Lambda}$
can be estimated using Equation~\eqref{eq:approxlambda}.

\begin{figure}[ht]
\includegraphics[width=0.5\textwidth]{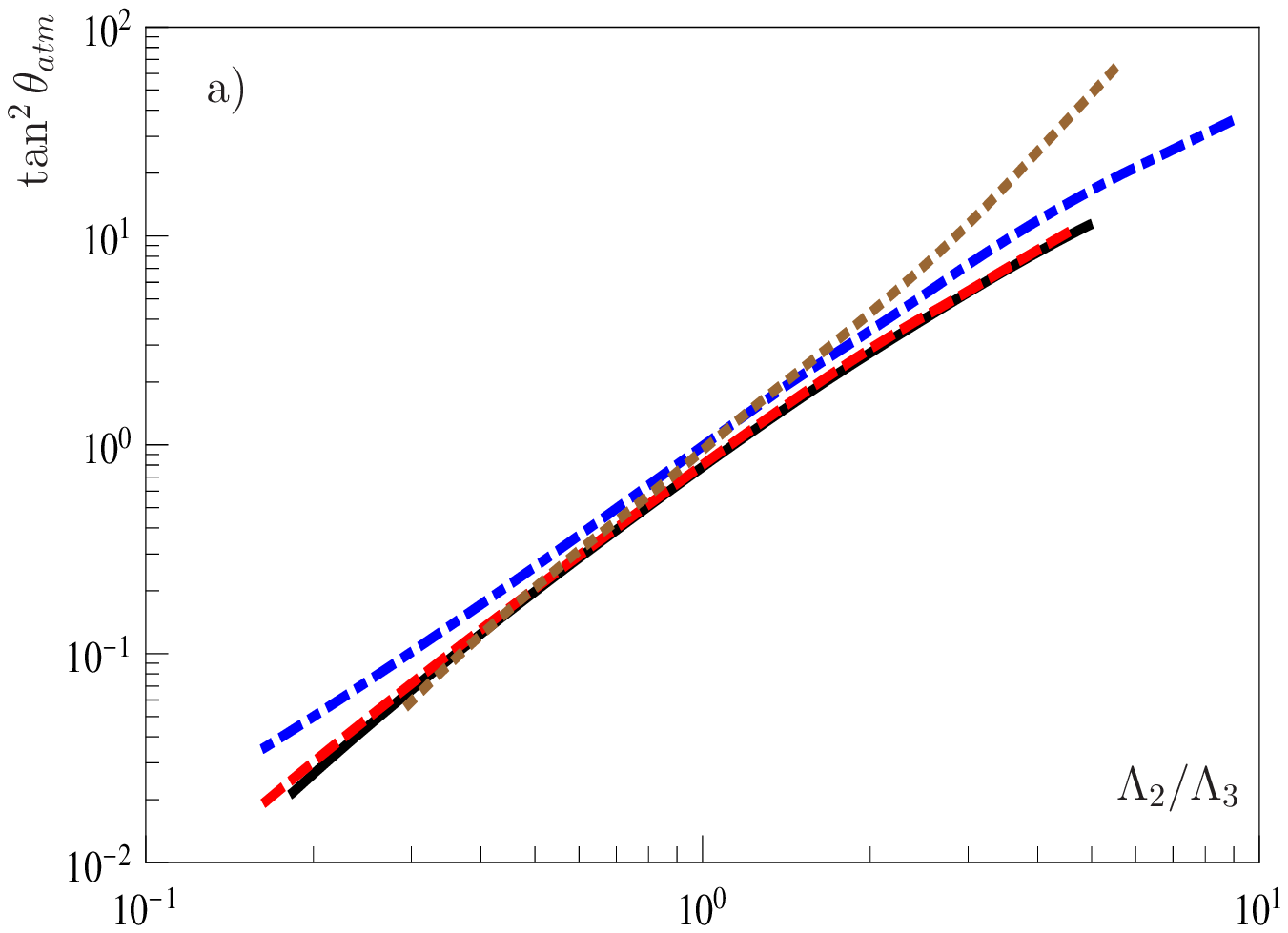}
\hfill
\includegraphics[width=0.5\textwidth]{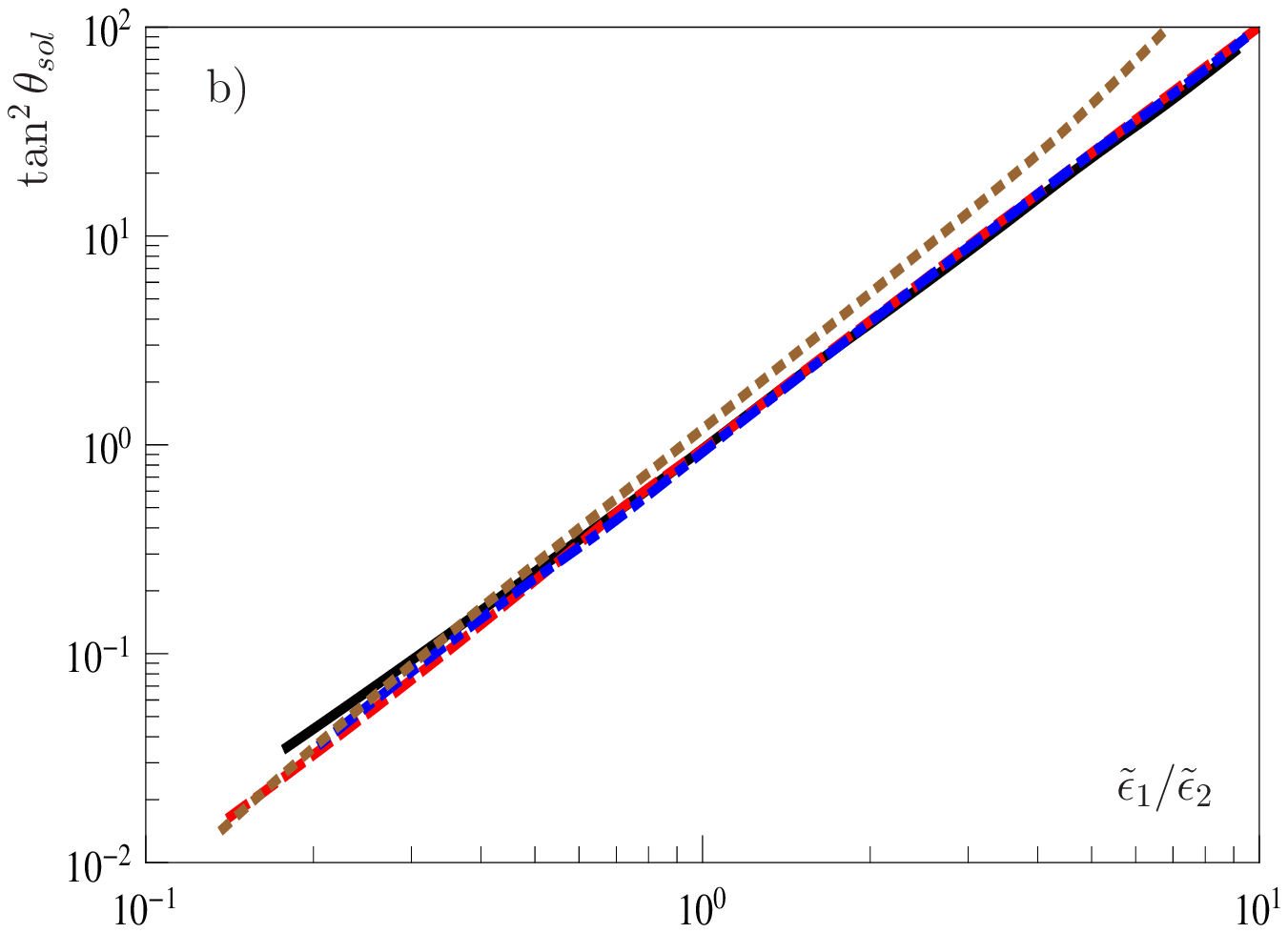}
\caption{Correlation between the alignment parameters and the neutrino
mixing angles for the $\mu\nu$SSM based on mSUGRA~$1$ (black),
$3'$ (red, dashed), $4'$ (blue, dot-dashed) and
GMSB~$5$ (brown, dotted) presented in Table~\ref{tab:4scmunuSSM}, in detail
a) $\tan^2\theta_{atm}$ as a function of $\Lambda_2/\Lambda_3$;
b) $\tan^2\theta_{sol}$ as a function of $\tilde{\epsilon}_1/\tilde{\epsilon}_2$.
}
\label{fig:Lambdaepstan2}
\vspace{-4mm}
\end{figure}

Using the definitions of Equation~\eqref{eq:mixingangles} for the neutrino mixing angles
we will now focus on the correlation between the alignment parameters
$\Lambda_i$ and $\tilde{\epsilon}_i$ and the neutrino mixing angles, which is shown
in Figure~\ref{fig:Lambdaepstan2}.
The ratio $\Lambda_2/\Lambda_3$ defines the atmospheric angle and
$\tilde{\epsilon}_1/\tilde{\epsilon}_2$ fits the solar angle.
The reactor angle is determined by $\Lambda_1/\sqrt{\Lambda_2^2+\Lambda_3^2}$
\cite{Hirsch:2000ef} which however can receive sizable loop corrections.

\section{Tree level decay widths of $\tilde{\chi}_1^0\rightarrow l^\pm W^\mp$}
\label{sec:treelevelwidth}

After having discussed the finite mass corrections between tree level and one-loop level,
which allow to explain the full neutrino spectrum, we will now turn to the
decays $\tilde{\chi}_1^0\rightarrow l^\pm W^\mp$ in order to study the correlations
between neutrino mixing angles and the different leptonic final states.
The partial widths for the decays
$\tilde{\chi}_j^{0}\rightarrow \tilde{\chi}_l^{\pm}W^{\mp}$ including
the \rpv decays $\tilde{\chi}_1^{0}\rightarrow l_l^{\pm}W^{\mp}$ are obtained
from the following interaction Lagrangian
{\allowdisplaybreaks\begin{align}
\mathcal{L}=\overline{\tilde{\chi}_l^-}\gamma^\mu\left(P_LO_{Llj}+P_RO_{Rlj}\right)\tilde{\chi}_j^0 W^-_\mu + h.c.\qquad.
\label{eq:treelevellagrangian}
\end{align}}\xspaceskip 0pt
The couplings are given by
{\allowdisplaybreaks\begin{align}
\label{eq:treelevelcouplings1}
O_{Llj} &= - g \Nge^*_{j2}U_{l1} - \frac{1}{\sqrt{2}} g \left(\Nge^*_{j3}U_{l2}+\sum_{k=1}^3 U_{l,2+k}\Nge_{j,4(5)+k}^*\right)\\
O_{Rlj} &= - gV_{l1}^*\Nge_{j2} + \frac{1}{\sqrt{2}} gV_{l2}^*\Nge_{j4} \quad ,
\label{eq:treelevelcouplings2}
\end{align}}\xspaceskip 0pt
where in Equation~\eqref{eq:treelevelcouplings1} the index is $4$ $(5)$ 
in case of  BRpV ($\mu\nu$SSM).
The widths have the form 
\begin{equation}
\Gamma^{0} = \frac{1}{16\pi m_i^3}\sqrt{\kappa(m_i^2,m_o^2,m_W^2)} \frac{1}{2}\sum_{pol}|M_T|^2\qquad,
\label{eq:leadingorderwidth}
\end{equation}
where $m_i$ ($m_o$) is the mass of the mother (daughter) particle and 
$M_T$ is the tree level matrix element. Explicitly they are given by
\begin{align}
\label{eq:neutdecaytree}
\Gamma^{\mathrm{0}}\left(\tilde{\chi}_j^{0}\rightarrow \tilde{\chi}_l^{+}W^{-}\right)
=& \frac{1}{16\pi m_j^3}\sqrt{\kappa(m_j^2,m_l^2,m_W^2)} \nonumber  \\ 
 \times \Big( \big(
|O_{Llj}|^2+&|O_{Rlj}|^2 \big) f(m_j^2,m_l^2,m_W^2)
-6 \mathrm{Re}(O_{Llj}O_{Rlj}^*)m_jm_l \Big) \\
\Gamma^{\mathrm{0}}\left(\tilde{\chi}_i^{+}\rightarrow \tilde{\chi}_k^{0}W^{+}\right)
=& \frac{1}{16\pi m_i^3}\sqrt{\kappa(m_i^2,m_k^2,m_W^2)} \nonumber  \\ 
 \times \Big( \big(
|O_{Lik}|^2+&|O_{Rik}|^2 \big) f(m_i^2,m_k^2,m_W^2)
-6 \mathrm{Re}(O_{Lik}O_{Rik}^*)m_im_k \Big) 
\label{eq:chardecaytree}
\end{align}
using the functions
\begin{align}
f(x,y,z) =& \frac{1}{2}(x+y)- z+\frac{(x-y)^2}{2 z} \\
\kappa(x,y,z)=&x^2+y^2+z^2-2xy-2xz-2yz\qquad.
\label{eq:kaellenfunction}
\end{align}

\subsection*{Coupling $\tilde{\chi}_1^0-l^{\mp}_i-W^{\pm}$}
\label{sec:AppCoupXiWl}

The tree level couplings $O_{Li1}$ in Equation~\eqref{eq:treelevelcouplings1}
and $O_{Ri1}$ in Equation~\eqref{eq:treelevelcouplings2} for the special case of the lightest
neutralino $\tilde{\chi}_1^0$ coupling to a lepton $l_i$ can be approximated in terms of the
alignment parameters $\Lambda_i$ and $\epsilon_i$, which
allows to understand the correlations to the neutrino mixing angles.
Although we have shown these approximated couplings in
great detail already in \cite{Bartl:2009an} for the $\mu\nu$SSM,
their knowledge is of great importance.
The right-handed couplings $O_{Ri1}$ can be approximated in BRpV and the
$\mu\nu$SSM by:
{\allowdisplaybreaks\begin{align}\label{Oapprox}
O_{Ri1} &\approx \frac{g (Y_l)_{ii}v_d}{2\mathrm{Det}_+}\left[ \frac{g v_u N_{12} - M_2 N_{14}}{\mu} \epsilon_i 
+\frac{g(2 \mu^2 +g^2 v_u^2)N_{12}-g^2 (v_d \mu +M_2v_u)N_{14}}{2 \mu \mathrm{Det}_+} \Lambda_i \right]
\end{align}}\xspaceskip 0pt
Therein the determinant of the heavy states in the chargino mass matrix appears
{\allowdisplaybreaks\begin{align}
\label{eq:det0}
\mathrm{Det}_+&=M_2\mu-\frac{1}{2}g^2v_dv_u\quad.
\end{align}}\xspaceskip 0pt
In contrast the left-handed couplings differ in both models and are given by the following expressions:
{\allowdisplaybreaks\begin{align}\nonumber
O_{Li1}^{b\text{\rpv}} \approx &\frac{g \Lambda_i}{\sqrt{2}}\left[ - \frac{g' M_2 \mu }{2 \mathrm{Det}_0^{b\text{\rpv}}} N_{11} 
  + \left( \frac{g}{\mathrm{Det}_+} + \frac{g M_1 \mu}{2 \mathrm{Det}_0^{b\text{\rpv}}} \right) N_{12} \right.\\
&\qquad - \frac{v_u}{2} \left( \frac{m_\gamma}{2 \mathrm{Det}_0^{b\text{\rpv}}}
  + \frac{g^2}{\mu \mathrm{Det}_+} \right) N_{13}
  + \left.\frac{v_d m_\gamma}{4 \mathrm{Det}_0^{b\text{\rpv}}} N_{14} \right] \\\nonumber
O_{Li1}^{\mu\nu\text{SSM}} \approx & \frac{g \Lambda_i}{\sqrt{2}}\left[
-\frac{g' M_2}{2 \mathrm{Det}_0^{\mu\nu\text{SSM}}}(v_d v_u \lambda^2 + m_R \mu)N_{11}\right.\\\nonumber
&\qquad
+\left(\frac{g}{\mathrm{Det}_+} + \frac{g M_1 }{2 \mathrm{Det}_0^{\mu\nu\text{SSM}}}
(v_d v_u \lambda^2 + m_R \mu)\right)N_{12}\\\nonumber
&\qquad-\left( \frac{g^2 v_u }{2 \mu \mathrm{Det}_+} 
       -\frac{m_\gamma}{8 \mu \mathrm{Det}_0^{\mu\nu\text{SSM}}}(\lambda^2 v_d (v_d^2 + v_u^2) + 2 m_R \mu v_u)\right)N_{13}\\\nonumber
&\qquad+\frac{m_\gamma}{8 \mu \mathrm{Det}_0^{\mu\nu\text{SSM}}}(\lambda^2 v_u (v_d^2 + v_u^2) + 2 m_R \mu v_d)N_{14}\\
&\qquad
\left.-\frac{\lambda m_\gamma}{4 \sqrt{2} \mathrm{Det}_0^{\mu\nu\text{SSM}}}(v_u^2 - v_d^2)N_{15}\right] 
 \end{align}}\xspaceskip 0pt
The couplings include the abbreviations of Equations~\eqref{eq:muepseff},\eqref{eq:lambdadefinition}
and \eqref{eq:abbreviationsapproxcoupl}
and the determinants shown in Equations~\eqref{eq:det1} and \eqref{eq:det2}.
Neglecting the right-handed couplings, which are generally smaller than
the left-handed ones, it is obvious that the lightest neutralino $\tilde{\chi}_1^0$
couples to $l_i^\pm W^\mp$ proportional to $\Lambda_i$ without any dependence on the $\epsilon_i$
parameters. This fact is in case of the considered BRpV and
the $\mu\nu$SSM independent of the nature of the lightest neutralino~$\tilde{\chi}_1^0$.
However, it is important to emphasize that all previous formulas are tree level results.
A priori it is not clear, what happens if one uses the mixing matrices $\Nge$, $U$
and $V$ for masses at one-loop level or performs a complete NLO calculation of
the decay width.

\section{NLO decay widths of $\tilde{\chi}_1^0\rightarrow l^\pm W^\mp$
and correlations to neutrino mixing angles}
\label{sec:oneloopwidth}

In this section we focus on the NLO decay widths
$\Gamma^1\left(\tilde{\chi}_1^0\rightarrow l^+W^-\right)$. The calculation
procedure was introduced in \cite{Liebler:2010bi} and can be adopted for
BRpV and the $\mu\nu$SSM without any changes.
The corresponding formulas have been implemented in the public program
\prog{} \cite{CNNDecaysonline}.
All technical questions concerning the used renormalization scheme like the cancellation
of UV and IR divergences or gauge invariance have been discussed in \cite{Liebler:2010bi}
for the (N)MSSM and preserve their validity for the $R$-parity violating extensions.

\subsection{Absolute size of the corrections}

\begin{figure}[ht]
\includegraphics[width=0.49\textwidth]{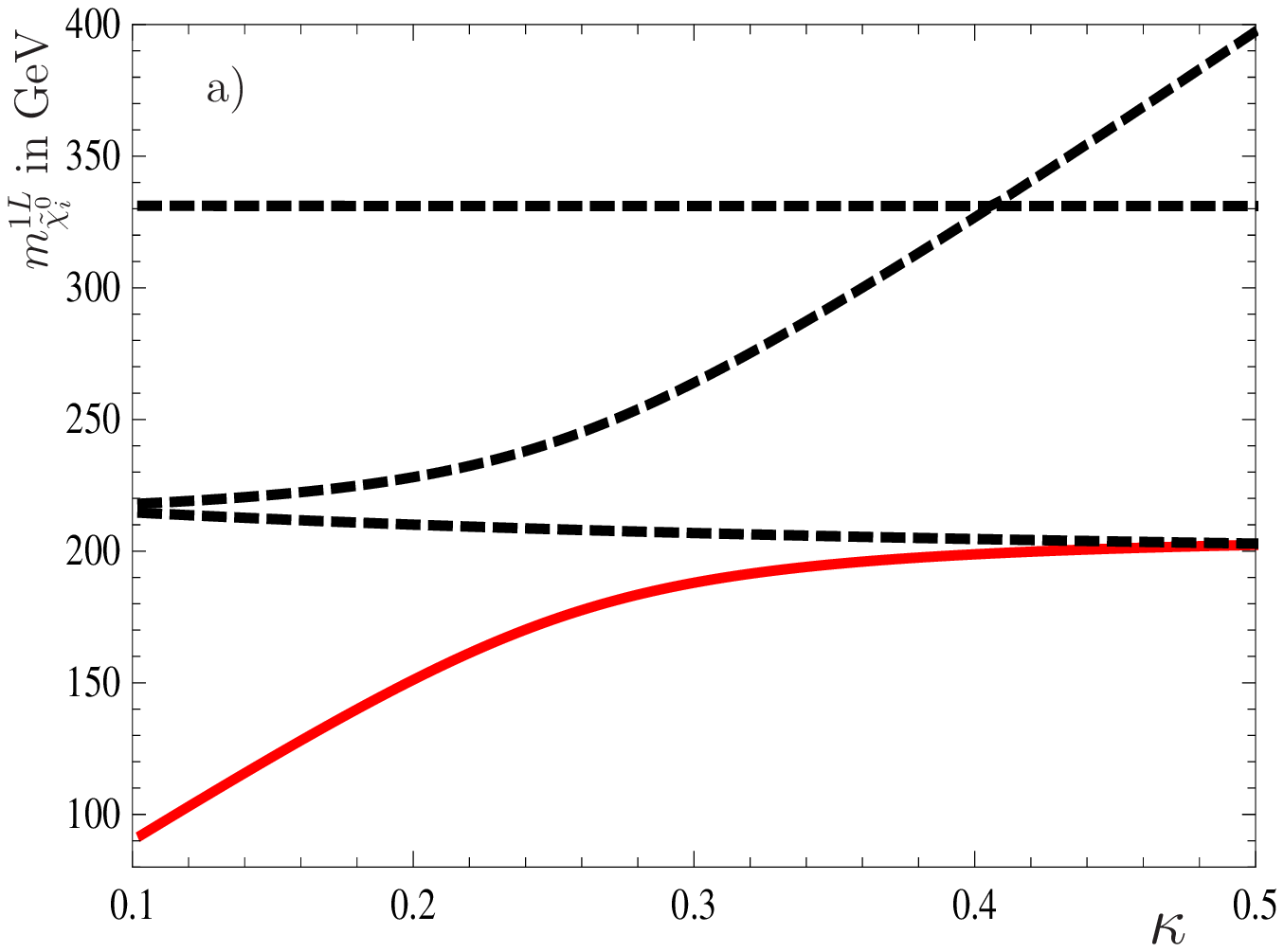}
\hfill
\includegraphics[width=0.49\textwidth]{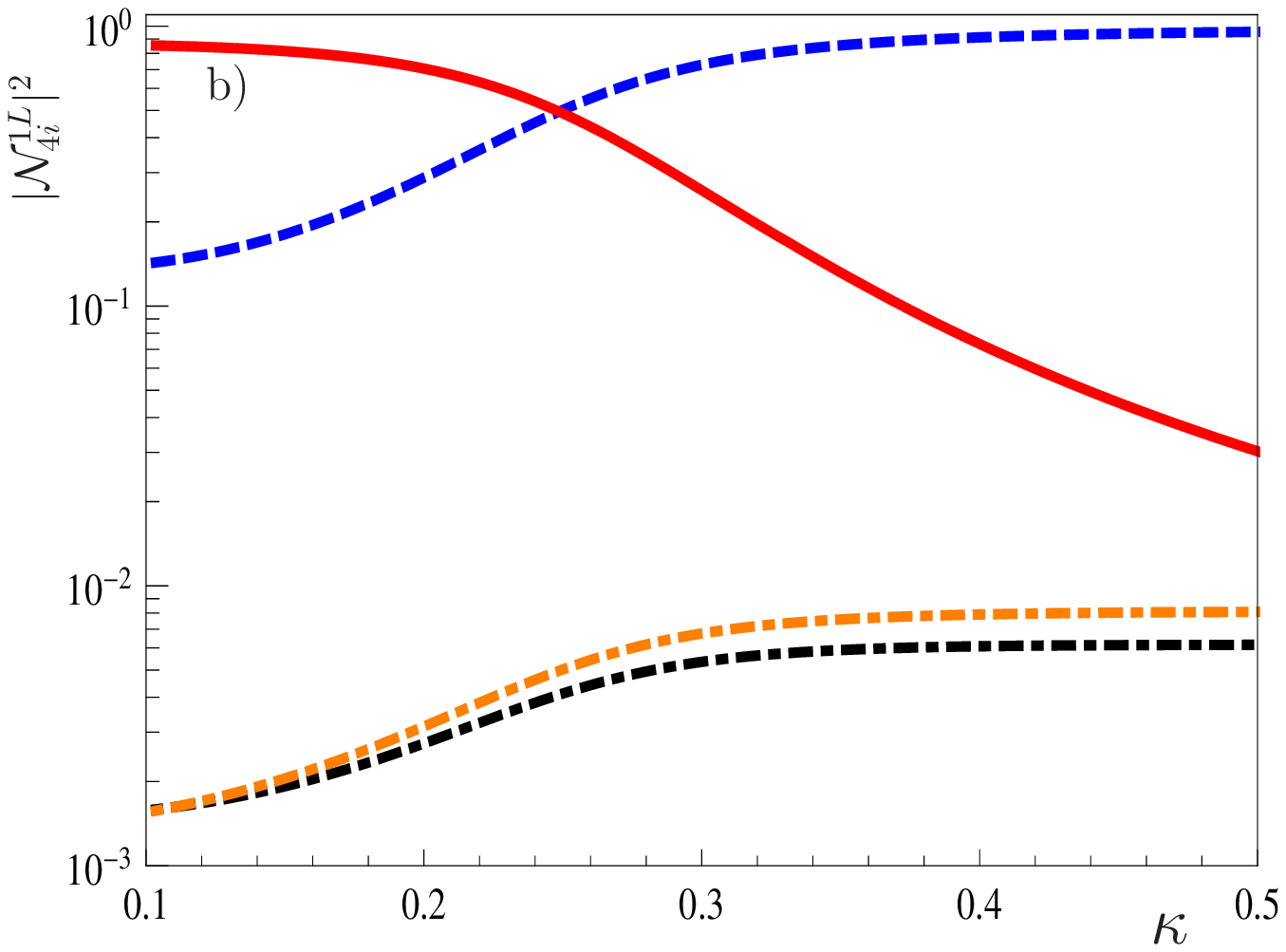}
\includegraphics[width=0.49\textwidth]{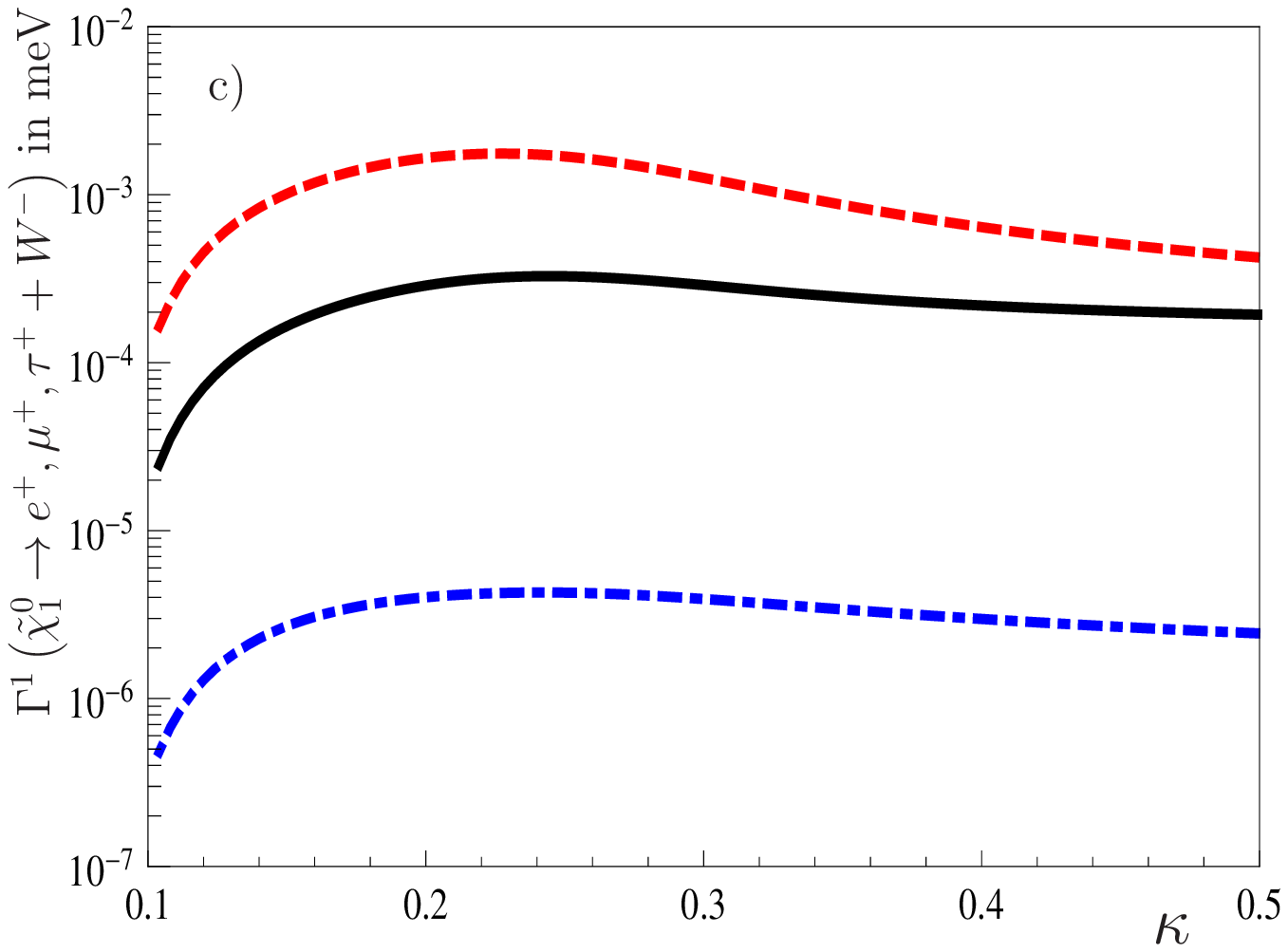}
\hfill
\includegraphics[width=0.49\textwidth]{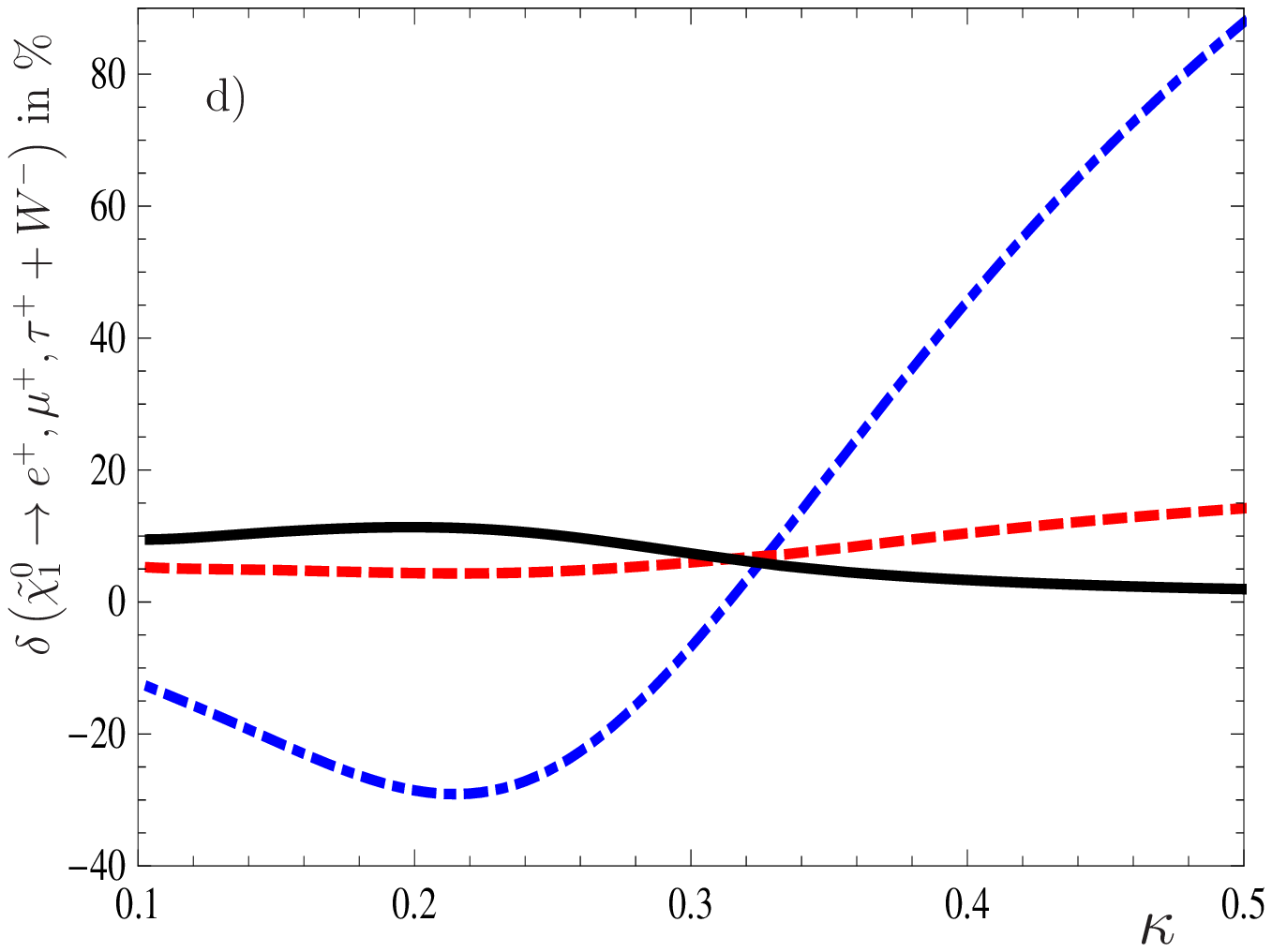}
\caption{Neutralino masses, particle content of the lightest neutralino 
and decay widths  $\Gamma^1(\tilde{\chi}_1^0\rightarrow l^+W^-)$
as a function of $\kappa$ within the $\mu\nu$SSM based on the mSUGRA~$4$ scenario and
$\vec \Lambda=(0.31,5.21,2.02)\cdot 10^{-2}$~GeV$^2$
and $\vec \epsilon=(7.49,9.61,-6.57)\cdot 10^{-3}$~GeV:
a) Neutralino masses $m_{\tilde{\chi}_i^0}^{1L}$ ($m_{\tilde{\chi}_1^0}^{1L}$ (red, solid),
$m_{\tilde{\chi}_{2,3,4}^0}^{1L}$ (black, dashed));
b) Particle character $|\Nge_{4i}^{1L}|^2$ 
($\tilde{\nu}^c$ (red, solid), $\tilde{H}_u+\tilde{H}_d$ (blue, dashed),
$\tilde{B}$ (black, dot-dashed), $\tilde{W}$ (orange, dot-dashed));
c) NLO decay width $\Gamma^1$ 
($e$ (blue, dot-dashed), $\mu$ (red, dashed), $\tau$ (black));
d) Relative correction $\delta$ defined in Equation~\eqref{eq:gammafactorwidth}
($e$ (blue, dot-dashed), $\mu$ (red, dashed), $\tau$ (black)).
}
\label{fig:msugra4kappavar}
\vspace{-4mm}
\end{figure}

In Figure~\ref{fig:msugra4kappavar} we present the NLO decay width for a
mSUGRA~$4$ based scenario, where we varied $\kappa=0.1-0.5$ in order
to have a singlino- and a higgsino-like lightest neutralino $\tilde{\chi}_1^0$
in comparison. For $\kappa\lesssim 0.1$ the decay is near the kinematical threshold and
for $\kappa\gg 0.5$ a Landau pole appears.
We show the relative correction $\delta (\tilde{\chi}_1^0\rightarrow l^+W^-)$,
which is defined by:
{\allowdisplaybreaks\begin{align}
 \delta (\tilde{\chi}_1^0\rightarrow l^+W^-) = \frac{\Gamma^1(\tilde{\chi}_1^0\rightarrow l^+W^-)
-\Gamma^0(\tilde{\chi}_1^0\rightarrow l^+W^-)}{\Gamma^0(\tilde{\chi}_1^0\rightarrow l^+W^-)}
\label{eq:gammafactorwidth}
\end{align}}\xspaceskip 0pt
The size and the sign of the corrections is strongly dependent on the
size and the sign of the $R$-parity violating parameters $v_{i}$ and $Y_{\nu}^i$
or $\Lambda_i$ and $\epsilon_i$.
In Figure~\ref{fig:msugra4kappavar} we set $\vec \Lambda=(0.31,5.21,2.02)\cdot 10^{-2}$~GeV$^2$
and $\vec \epsilon=(7.49,9.61,-6.57)\cdot 10^{-3}$~GeV.
Since the solar mass difference and mixing angle are only well
defined at  one-loop level, the corrections to the decay
 $\tilde{\chi}_1^0\rightarrow e^+W^-$
are sizable and can be of the order of the tree level decay width or larger.
The corrections to $\tilde{\chi}_1^0\rightarrow \mu^+W^-$ or $\tau^+W^-$
are generally smaller, but remain  of order $10\%$.

\subsection{Correlation between decay widths and neutrino mixing angles}

\begin{figure}[t]
\includegraphics[width=0.5\textwidth]{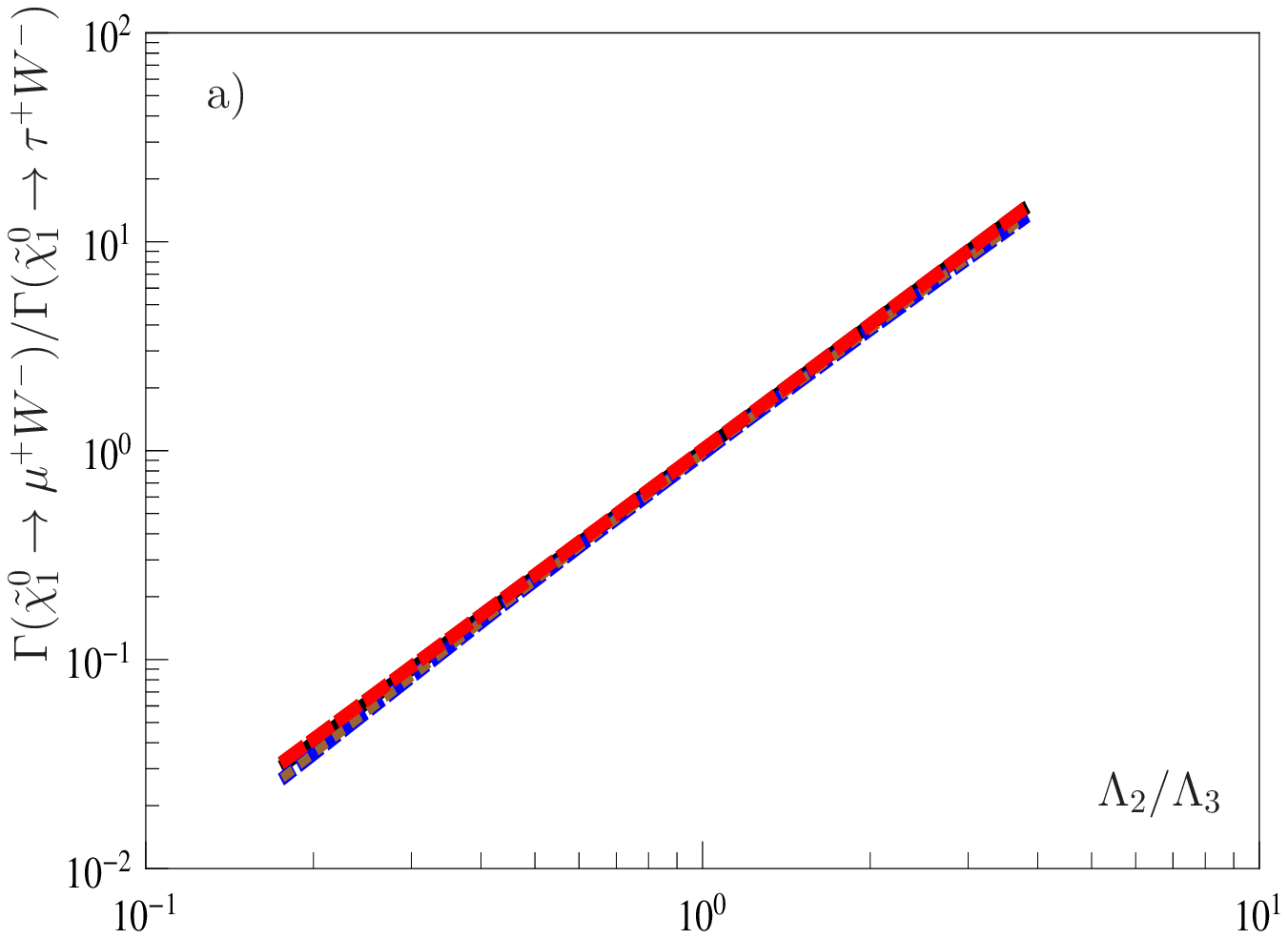}
\hfill
\includegraphics[width=0.5\textwidth]{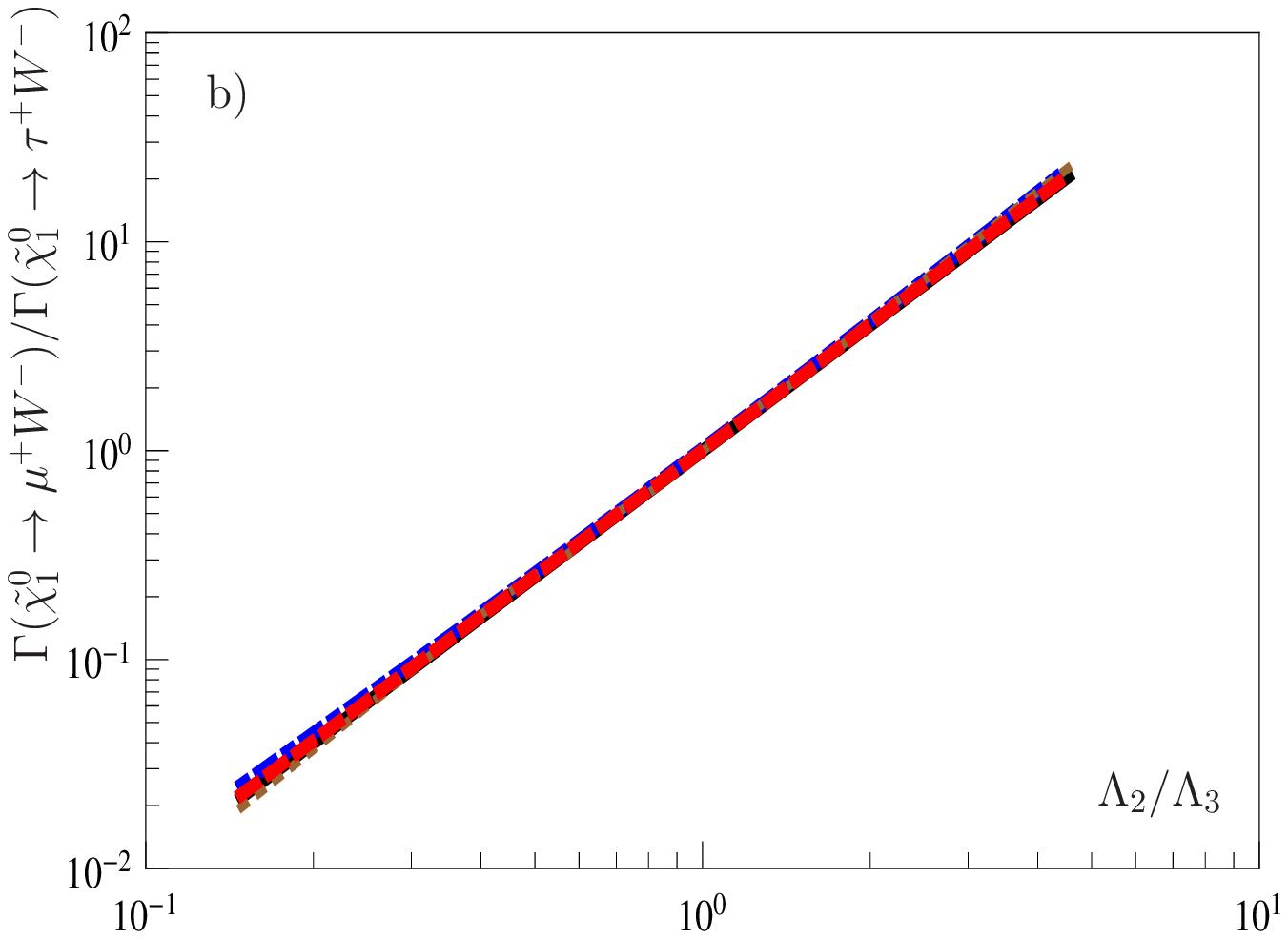}
\includegraphics[width=0.5\textwidth]{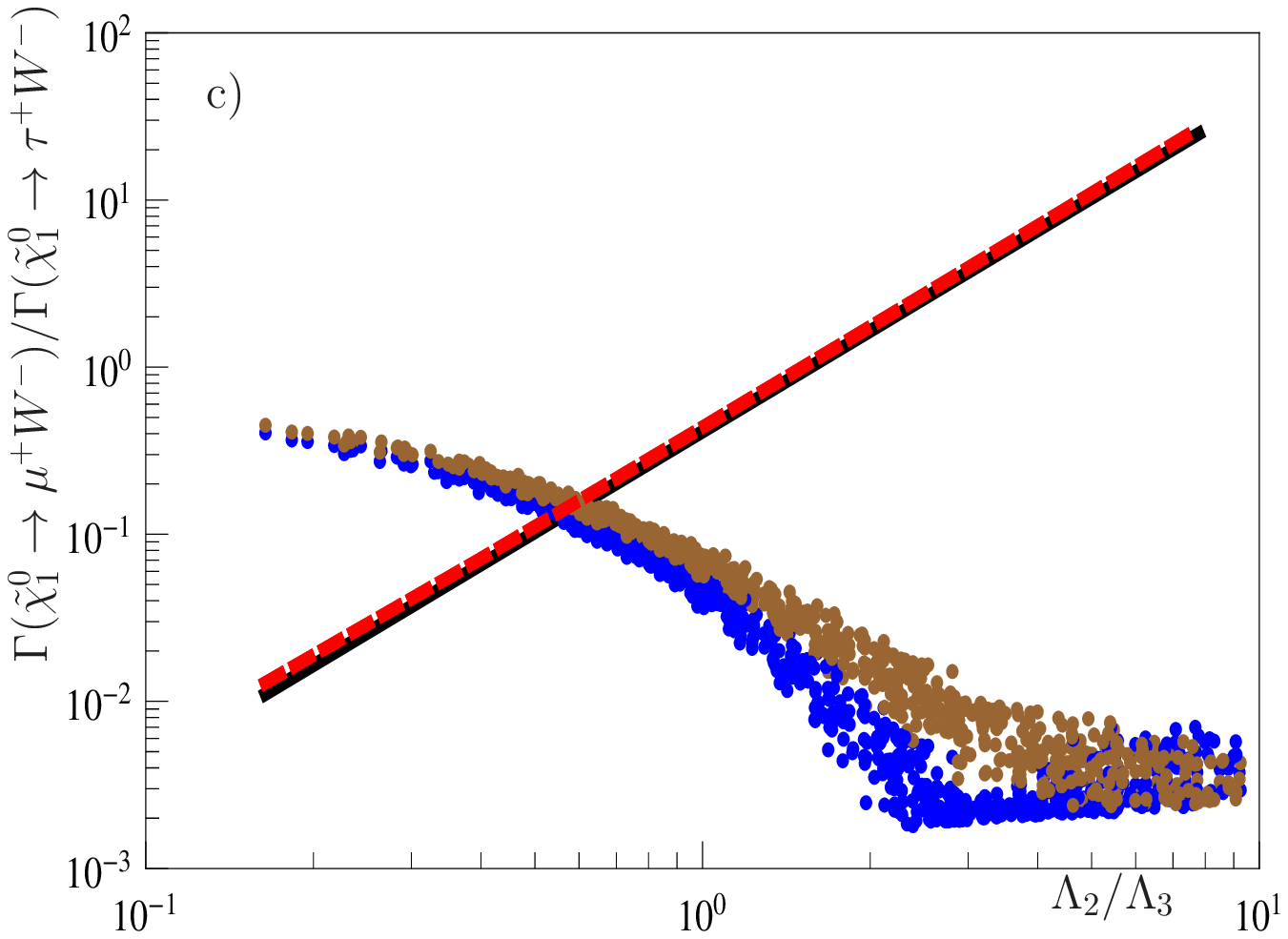}
\hfill
\includegraphics[width=0.5\textwidth]{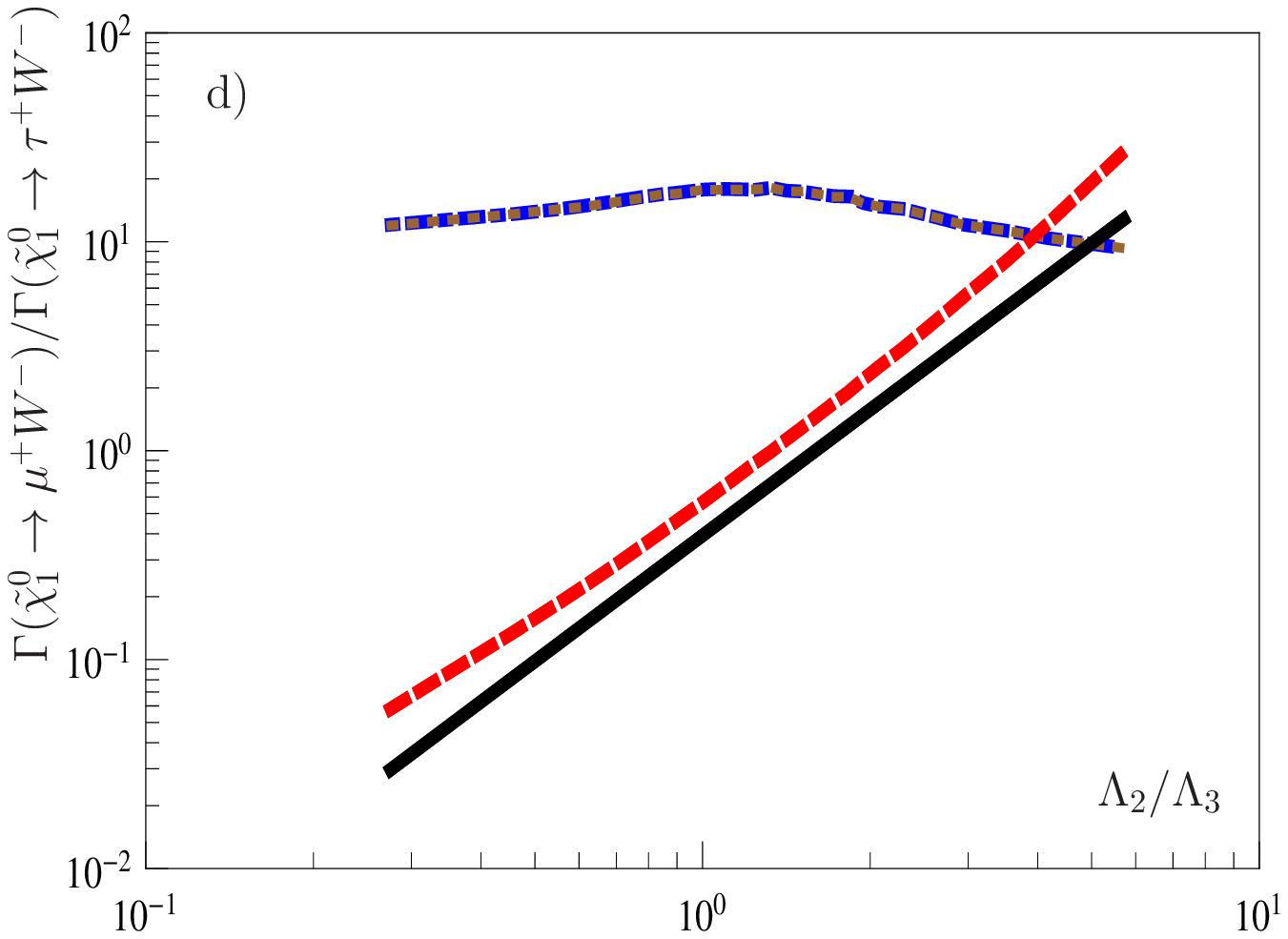}
\caption{
Ratios $\Gamma(\tilde{\chi}_1^0\rightarrow \mu^+W^-)/\Gamma(\tilde{\chi}_1^0\rightarrow \tau^+W^-)$
with $\Gamma=\Gamma^0$ in black, $\Gamma=\Gamma^1$ in red, dashed, $\Gamma=\Gamma^0$ with $\Nge^{1L}$ in blue, dot-dashed,
$\Gamma=\Gamma^0$ with $\Nge^{1L},U^{1L}$ and $V^{1L}$ in brown, dotted as a function of $\Lambda_2/\Lambda_3$
for the $\mu\nu$SSM scenarios of Table~\ref{tab:4scmunuSSM}:
a) mSUGRA~$1$;
b) mSUGRA~$3'$;
c) mSUGRA~$4'$;
d) GMSB~$5$.
}
\label{fig:relationfor4sc}
\vspace{-4mm}
\end{figure}

Having discussed the absolute size of the NLO contributions to the decay widths of the
two body decays $\tilde{\chi}_1^0\rightarrow l^+W^-$, we will now focus on
the correlations between the branching ratios and the neutrino mixing angles.
We have shown in Section~\ref{sec:AppCoupXiWl}, that the tree level couplings
can be approximated in terms of the alignment parameters $\vec{\Lambda}$ and $\vec{\epsilon}$.
Since the alignment parameters are connected to the neutrino mixing angles
as explained in Section~\ref{sec:rellambdaepsangles},
one can expect the following tree level relation:
{\allowdisplaybreaks\begin{align}
\frac{\Gamma^0(\tilde{\chi}_1^0\rightarrow \mu^+W^-)}{\Gamma^0(\tilde{\chi}_1^0\rightarrow \tau^+W^-)}
\propto \left(\frac{O_{L21}}{O_{L31}}\right)^2\approx \left(\frac{\Lambda_2}{\Lambda_3}\right)^2\approx\tan^2\theta_{atm}
\end{align}}\xspaceskip 0pt
It is a priori not clear to which extent this relation also holds
at one-loop level. We have found that using one-loop corrected
$\overline{\text{DR}}$ masses together with the corresponding one-loop corrected
mixing matrix $\Nge^{1L}$ in the tree level decay width can spoil this
relation. However, performing the full one-loop corrections including
the real corrections from photon emission in the considered on-shell scheme
retains the predictions at tree level, so that
{\allowdisplaybreaks\begin{align}
\frac{\Gamma^1(\tilde{\chi}_1^0\rightarrow \mu^+W^-)}{\Gamma^1(\tilde{\chi}_1^0\rightarrow \tau^+W^-)}
\propto  \left(\frac{\Lambda_2}{\Lambda_3}\right)^2\approx\tan^2\theta_{atm}\quad.
\end{align}}\xspaceskip 0pt
In Figure~\ref{fig:relationfor4sc} we highlight this feature using the 
scenarios of Table~\ref{tab:4scmunuSSM} for the $\mu\nu$SSM. We do not
only show the tree level and one-loop relations,
but also the relations obtained by using
the one-loop mixing matrices $\Nge^{1L},U^{1L}$ and $V^{1L}$ within the
tree level decay width $\Gamma^0(\tilde{\chi}_1^0\rightarrow l^+W^-)$,
see Equation~\eqref{eq:neutdecaytree}. The ratios are shown as a function of
$\Lambda_2/\Lambda_3$, which is equivalent to varying $\tan^2\theta_{atm}$
as can be seen in Figure~\ref{fig:Lambdaepstan2}. We vary the other
parameters such, that the neutrino data are within their $2\sigma$ bounds
as given in Table~\ref{tab:neutrinobounds}. This is the cause of the observed
widths in case of the mSUGRA~$4'$ scenario.

As already known from BRpV the approximated one-loop contributions
using $\Gamma^0$ together with the one-loop corrected mixing matrices show
the expected behavior if the lightest neutralino is either mainly
a bino, wino or a pure higgsino. However, in case that
the singlino-component gets sizable or dominant,
 these approximations fail and the complete one-loop decay width
$\Gamma^1$ has to be calculated to obtain a reliable result.
In case of mSUGRA~$4'$, 
which is shown in Figure~\ref{fig:relationfor4sc},
the higgsino has a purity of $91.7\%$, but a singlino component of
$6.9\%$ is left, so that also in this case the full one-loop decay
width is advisable.

The reason for the misleading ratios in case of a singlino-like neutralino
is a cancellation between the second and the last terms of the
left-handed coupling $O_{Ll1}$ in Equation~\eqref{eq:treelevelcouplings1} at tree level.
Replacing the tree level mixing
matrix $\Nge$ by the one-loop mixing matrix $\Nge^{1L}$ spoils
this cancellation, so that unreasonable results for the decay widths
and also the ratios are obtained. However, the very same effect is 
restored once the complete one-loop corrections are included.

\begin{figure}[ht]
\includegraphics[width=0.5\textwidth]{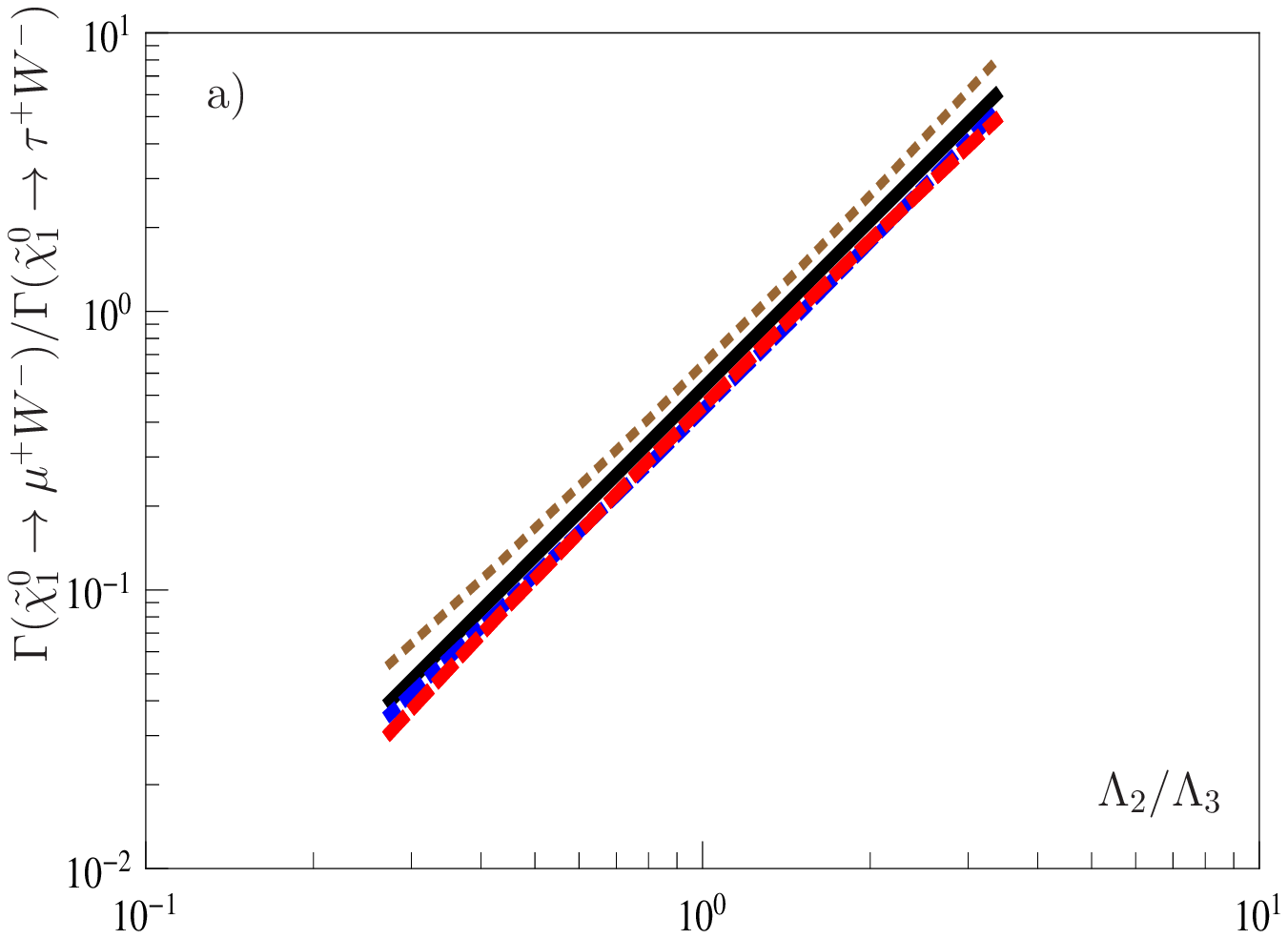}
\hfill
\includegraphics[width=0.5\textwidth]{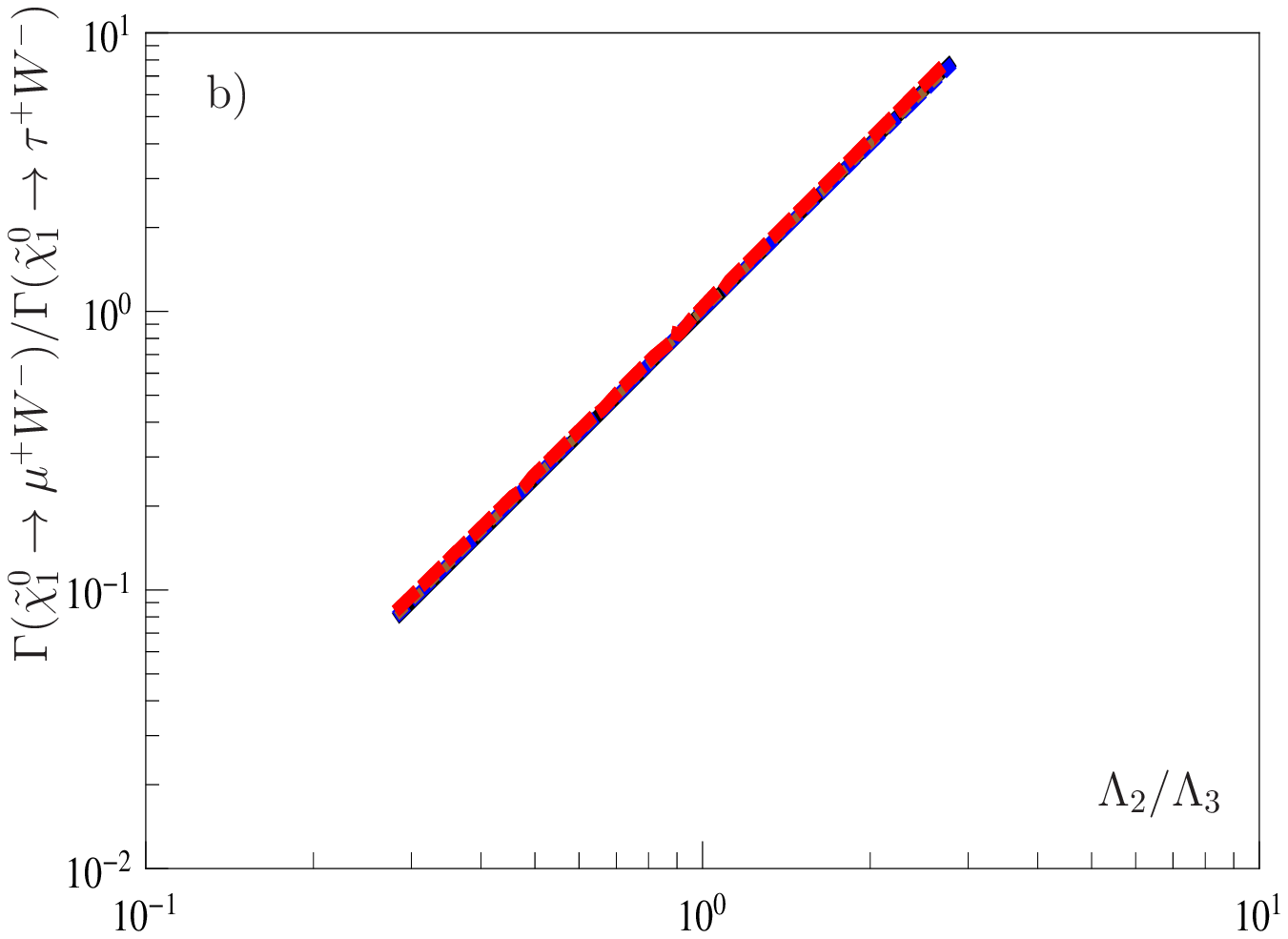}
\caption{
Ratios $\Gamma(\tilde{\chi}_1^0\rightarrow \mu^+W^-)/\Gamma(\tilde{\chi}_1^0\rightarrow \tau^+W^-)$
with $\Gamma=\Gamma^0$ in black, $\Gamma=\Gamma^1$ in red, dashed, $\Gamma=\Gamma^0$ with $\Nge^{1L}$ in blue, dot-dashed,
$\Gamma=\Gamma^0$ with $\Nge^{1L},U^{1L}$ and $V^{1L}$ in brown, dotted as a function of $\Lambda_2/\Lambda_3$
for the BRpV scenarios of Table~\ref{tab:4scmunuSSM}:
a) SPS~$2'$;
b) SPS~$3$.
}
\label{fig:relationfor2sc}
\vspace{-4mm}
\end{figure}

For completeness let us briefly discuss the situation within BRpV.
As examples we take scenarios  where the lightest neutralino is either mainly
a bino, scenario SPS~$3$, or mainly a higgsino, scenario SPS~$2'$.
The ratios of decay widths using the various approximations are shown in
Figure~\ref{fig:relationfor2sc}. We observe that NLO corrections
are somewhat more important for higgsino-like neutralinos:
Here the combination of tree level decay widths with NLO masses
and mixing matrices can lead to correlations which are off by
a factor 2 compared the case of the complete NLO calculation.
Moreover, the complete NLO corrections can shift the correlations
up to 30 per-cent compared to the tree level result.

\newpage
\section{Conclusion}
\label{sec:conclusion}

In \rpv models, where the sneutrinos obtain a VEV, a mixing between
charged leptons and charginos as well as a mixing between
neutrinos and neutralinos occur at tree level. 
The latter can be used to obtain at least one neutrino mass at tree level.
The other neutrino masses can be generated at the one-loop level as it
happens
e.g.\ in bilinear \rpv or in the $\mu\nu$SSM with one generation of
right-handed neutrinos. In previous work a $\overline{\text{DR}}$ scheme
had been used for this. In this paper we worked out an on-shell scheme
and compared both schemes. The challenge here is, that there are less
parameters than masses implying finite shifts at the one-loop level to
the masses.
With respect to the masses of neutrinos
and charged leptons we have found only small differences well below
the current and near future experimental accuracies.

In both models one expects correlations between ratios of the
decay widths or equivalently the branching ratios for the decays
$\tilde{\chi}^0_1 \rightarrow l^\pm W^\mp $ and the atmospheric mixing angle
which can be shown easily at tree level using approximate formulas for
the mixing matrices. We have calculated these decays including
the complete NLO order corrections in the on-shell scheme and
compared the results to the pure tree level as well as to the
case where one uses tree level formulas for the decay widths but
one-loop corrected masses and mixing matrices.

We have found that corrections to the channels containing either
a $\mu$ or a $\tau$ lepton are typically of order ten per-cent
whereas the corrections for the $e$-channel can be of the order
of the tree level decay width.
The reason for the latter finding is that the solar mixing angle is
not well defined at tree level but only at the one-loop level.
For the correlations between the decay widths and the atmospheric
mixing angle we find that in case of gaugino-like neutralino in 
all approximations nearly the same correlations. However, already
for doublet-higgsino-like neutralino the correlations can be off
by a factor two if one combines tree level widths with one-loop
corrected masses and mixing angles. In case of a neutralino
which has either a sizable singlino-component or which is
dominantly singlino-like
this is even more pronounced and the correlations
get completely wrong in this approximation. Here one has to use
the complete NLO corrections to obtain a reliable prediction.
Last but not least we note that both models have
been implemented in 
the publicly available program \prog{} for the numerical evaluation.

\section{Acknowledgments}

This work has been supported by the DFG, Project no. PO-1337/2-1. 
S.L.\ has been supported by the
DFG research training group GRK1147.

\appendix

\section{New blocks in \prog}
\label{sec:CNNDecaysapp}

Both considered models, BRpV and the $\mu\nu$SSM with
one right-handed neutrino superfield, are now supported by \prog,
which was published in \cite{Liebler:2010bi} and can be found
online \cite{CNNDecaysonline}. We provide input files
for the MSSM, the NMSSM, BRpV and the $\mu\nu$SSM with
one right-handed neutrino superfield in the folder
{\tt /examples}, where in addition the corresponding output
files can be found. In BRpV $\epsilon_i$ and $v_i$ can
be chosen beside the input parameters of the MSSM:
\lstset{basicstyle=\scriptsize, frame=shadowbox}
\begin{lstlisting}
Block MODSEL                 # Select model
 1    0                      # bilinear model
 4    1			     # RPviolation
....
BLOCK RVKAPPAIN
     1    1.45660382E-02    # kappa_1 = eps_1
     2    9.01765562E-03    # kappa_2 = eps_2
     3   -3.16131217E-03    # kappa_3 = eps_3
BLOCK RVSNVEVIN
     1   -8.68089903E-04    # v_L_1
     2   -4.50162251E-04    # v_L_2
     3    4.19592513E-04    # v_L_3
....
\end{lstlisting}

The input file of the $\mu\nu$SSM follows the convention
of the NMSSM, implying that the VEV of the right-handed sneutrino $v_c$
is calculated from the effective $\mu=\tfrac{1}{\sqrt{2}}\lambda v_c$.
In addition $\kappa$ and the soft-breaking terms $A_\lambda$ and $A_\kappa$ are input.
The $R$-parity violating parameters should be provided
via the variables $Y_\nu^i$ and $v_i$:
\begin{lstlisting}
Block MODSEL                 # Select model
 3    6                      # munuSSM
 4    1			     # RPviolation
....
BLOCK EXTPAR
....
    61    1.00000000E-01    # lambda
    62    1.08485437E-01    # kappa/2
    63   -9.59966990E+02    # T_lambda/lambda = A_lambda
    64   -1.58051889E+00    # T_kappa/kappa = A_kappa
    65    9.68523016E+02    # mu_eff
    73    5.71920838E-06    # Y_nu_1
    74    6.20206893E-06    # Y_nu_2
    75   -6.20206893E-06    # Y_nu_3
BLOCK RVSNVEVIN
     1   -1.35445088E-03    # v_L_1
     2   -1.27271724E-03    # v_L_2
     3    1.71364110E-03    # v_L_3
....
\end{lstlisting} 
The corresponding soft-breaking terms of $Y_\nu^i$, namely
$T_\nu^i$, are calculated from the tadpole equations in addition
to $m_{H_d}^2$, $m_{H_u}^2$ and ${m_{\tilde{\nu}^c}}^2$.
The output does not only contain the $R$-parity violating decays
of the lightest neutralino $\tilde{\chi}_1^0\rightarrow l^\pm W^\mp$,
but in addition the relevant parameters for neutrino physics:
\begin{lstlisting}
....
DECAYTREE   1000022     2.03988581E-13   # chi01
#    BR                NDA   ID1       ID2
     5.00974421E-01    2          15       -24   # BR(chi01 -> tau+ W-)
     4.94172793E-01    2          13       -24   # BR(chi01 -> mu + W-)
     4.85278632E-03    2          11       -24   # BR(chi01 -> e+ W-)
DECAY   1000022     2.17956335E-13   # chi01 on NLO
#    BR                NDA   ID1       ID2
     4.99329442E-01    2          15       -24   # BR(chi01 -> tau+ W-)
     4.95513455E-01    2          13       -24   # BR(chi01 -> mu + W-)
     5.15710289E-03    2          11       -24   # BR(chi01 -> e+ W-)
....
Block SPhenoRP  # additional RP parameters
         1     5.53918495E-02  # eps_1
         2     6.00684651E-02  # eps_2
         3    -6.00684651E-02  # eps_3
         4     2.10857140E-02  # Lambda_1 = v_d epsilon_1 + mu v_L1 [GeV^2]
         5     2.12780743E-01  # Lambda_2 = v_d epsilon_2 + mu v_L2 [GeV^2]
         6     2.14264164E-01  # Lambda_3 = v_d epsilon_3 + mu v_L3 [GeV^2]
         7     2.24691592E-03  # m^2_atm [eV^2]
         8     7.74812242E-05  # m^2_sol [eV^2]
         9     9.27311984E-01  # tan^2 theta_atm
        10     4.08412270E-01  # tan^2 theta_sol
        11     3.54098593E-03  # U_e3^2
        15     2.40631533E+01  # v_d
        16     2.40631533E+02  # v_u
        17    -1.35445088E-03  # v_L1
        18    -1.27271724E-03  # v_L2
        19     1.71364110E-03  # v_L3
\end{lstlisting}
In case of the $\mu\nu$SSM the given $\epsilon_i$ are calculated
from Equation~\eqref{eq:muepseff}. The screen output moreover contains not
only the one-loop on-shell neutralino and chargino masses, but
also the tree level masses for comparison.
Please note that for the correct calculation of the one-loop
on-shell lepton masses a high accuracy of more than $20$ digits is needed
to obtain numerical stable results.
In most cases for neutrino masses and the decay widths $15$ digits
are sufficient, as long as all parameters are chosen real, implying
that the diagonalization process isn't based on the
squared mass matrices. To allow for a higher accuracy, adopt
Selected\_real\_kind$(15,300)$ to $(25,300)$ in {\tt /sphenooriginal/Control.F90}
and compile \prog. We note that a higher accuracy
can result in significantly longer  computing times.

\bibliographystyle{h-physrev}

\end{document}